\documentclass[11pt,a4paper]{article}
\usepackage{jcappub}
\usepackage{graphicx}
\usepackage{subfigure}
\usepackage{caption}

\title{Cosmological constraints on sterile neutrino Dark Matter production mechanisms}

\author{Lucia Aurelia Popa}

\affiliation{Institute of Space Science,\\
Bucharest-Magurele, Ro-077125 Romania}

\emailAdd{lpopa@spacescience.ro}

\abstract{ We place constraints on sterile neutrino resonant production (RP) and scalar decay production (SDP)  mechanisms  assuming
that sterile neutrino represents a fraction $f_S$ from the total Cold Dark Matter energy density. \\
For the cosmological analysis, we complement the CMB anisotropies measurements with CMB lensing gravitational potential measurements, that are
sensitive to the DM  distribution  out to high redshifts
and with the cosmic shear data, that constraints the gravitational potential at lower redshifts than CMB. We also use the most recent low-redshift BAO measurements that are insensitive  to the non-linear effects,
providing robust geometrical tests.\\
We show that our datasets have enough sensitivity to constrain 
the sterile neutrino mass and mass fraction inside the co-moving free-streaming 
horizon in both RP and SDP scenarios. \\
For RP case we find that the best fit values of sterile neutrino mass and mixing angle 
are in the parameter space of interest for sterile neutrino DM decay interpretation of the 3.5 keV X-ray line
with a DM mass fraction $f_{S}=0.28 \pm 0.3$  (at 68\% CL) that excludes  the assumption of
sterile neutrinos as being all of the DM.  For SDP case we find $f_{S}=0.86 \pm 0.07$  (at 68\% CL), 
in agreement with the upper limit constraint on $f_S$ from the X-ray non-detection and Ly-$\alpha$  
forest measurements that rejects $f_S=1$  at 3$\, \sigma$ level \cite{Slosar}. \\
The sterile neutrino mass predicted by both RP and SDP models are consistent within 0.3$\sigma$.\\
We analysed the possibility to distinguish  between  RP and SDP scenarios through their impact on the
acoustic scales, the small scale fluctuations and the low-redshift geometric observables, obtaining 
cosmological constrains that clearly show that  
the present-day cosmological data start to discriminate between different 
sterile neutrino DM production mechanisms. \\
However, we expect 
the future BAO and weak lensing surveys, such as
{\sc EUCLID}, to provide much robust constraints.}

\def\be{\begin{equation}}
\def\ee{\end{equation}}
\def\ba{\begin{eqnarray}}
\def\ea{\end{eqnarray}}

\begin{document}
\begin{flushright}
\end{flushright}

\maketitle
\section{Introduction} 

\hspace{0.7cm} Cosmic Microwave Background (CMB) measurements from the {\sc Planck} satellite, alone or in combination with other astrophysical datasets, provide no powerful evidence supporting new physics  beyond the standard $\Lambda$CDM cosmological model \cite{Planck13,Planck16,Planck18}. \\
With around 5\% of the total energy density of the universe representing the baryonic matter, 21\% the Dark Matter (DM) and 74\% accounting for the Dark Energy (DE), the $\Lambda$CDM model is remarkably successful at reproducing the large-scale structure (LSS) of the universe. 
In addition, the Planck results show that the signature of neutrino sector is consistent with the $\Lambda$CDM model assumptions and that DE is compatible with  the $\Lambda$ cosmological constant. \\
Some tension still exists between the {\sc Planck} determination of several observables and their values obtained from astrophysical independent probes. The most notable tension concern the smaller value of the Hubble constant, $H_0$, discordant at about $2.5 \sigma$ level with the value obtained from direct astrophysical measurements \cite{HST,CHP,RXJ}. Also,  {\sc Planck} determination of $\sigma_{8} $  (the amplitude of linear power spectrum on scale of $8h^{-1}$ Mpc, $h$ being 
the reduced Hubble constant, $h=H_0/(100$ km s$^{-1}$  Mpc$^{-1}$) and of matter energy density, $\Omega_m$, are discordant at $2 \sigma$ level with  the corresponding values inferred from cluster data that prefer lower values of these observables \cite{s8-om1,s8-om2}. 
These discrepancies  may arise  because of biases and calibration errors of direct astrophysical measurements \cite{Planck16,Marra}
but may also be related to the assumption of the underlying $\Lambda$CDM cosmological model \cite{Reconciling}. 

%-----------------DE-----------------------
Interpretation of DE in the form of $\Lambda$ is facing challenges  such as the cosmological constant problem \cite{lambda} and the coincidence problem \cite{coincidence}. The first problem refers to the small observed value of $\Lambda$, incompatible with the prediction of the field theory. The second problem regards the fact that there is not a natural explanation why DM and DE energy densities are of the same order of magnitude today. 
Alternative DE models aiming to alleviate these problems have been proposed. In these models DE is generally described by a dynamical cosmological fluid associated either to a scalar field  \cite{quint2} or to modifications of gravity \cite{grav1,grav2}, 
although a quantum running of $\Lambda$ could  provide a satisfactory evolving DE scenario \cite{quint1,shapiro,sola}. 
%---------DM-------------------------------------------

The nature and composition of DM is still unknown. Attempts involving collision-less DM 
particles fail to solve the $\Lambda$CDM problems  at reproducing the cosmological 
structures at small scales  (missing satellite problem \cite{missing1,missing2,missing3}, 
core-cusp problem  \cite{crusp1,crusp2,crusp3},  too-big-to-fail problem \cite{fail1,fail2} ), 
suggesting that DM particles may also exhibit gravitational properties and 
requiring the  extension of the Standard Model (SM) of particle physics \cite{ext1,ext2,ext3}. \\
The Weakly Interacting Massive Particles (WIMPs) with masses above the electroweak scale 
are good DM candidates \cite{wimp1}. As WIMPs decouple from the thermal plasma when the Hubble expansion rate becomes larger than their interaction rate (thermal {\it freeze-out}).
Although well theoretically  motivated, currently no conclusive WIMPs experimental evidences have been found (see e.g. \cite{snwp} and references therein). \\
Another theoretically well motivated DM candidate is sterile neutrino \cite{sn1,sn2,sn3,sn4}. 
Arising in the minimal extension of SM, the sterile neutrino with mass in  keV range
can simultaneously explain the active neutrino oscillations, the DM properties and the  
matter-antimatter asymmetry of the universe \cite{sdm1,sdm2}. Detection of a weak X-ray emission line at an energy of $\sim$3.5 keV from clusters and 
Andromeda galaxy  independently reported by XMM-Newton and Chandra satellites \cite{x1,x2} initiated a large debate 
on the possibility that this line is the signature of  DM decay \cite{xc1,xc2,xc3}. If confirmed,   
this signal could be the signature of decaying sterile neutrino DM with a mass of 7.1 keV \cite{line}. \\
As sterile neutrinos are weakly interacting particles they cannot be produced 
in the early universe by {\it thermal freeze-out}. 
Instead they could be gradually produced from the thermal plasma by the  {\it thermal  freeze-in}  \cite{freeze-in} with non-thermal spectrum, the dominant production occurring  when the temperature drops below the sterile neutrino mass. 
Several  keV sterile neutrino DM production mechanisms  have been proposed.\\
In the Dodelson-Widrow (DW) scenario \cite{DW}, keV sterile neutrinos DM are produced by
non-resonant oscillations  with active neutrinos in presence of negligible leptonic asymmetry. 
This mechanism is  now excluded by the observations of structure formation as it produces too 
hot sterile neutrino velocity spectra \cite{NRP1,NRP2}. \\
The keV sterile neutrino DM resonant production (RP) via the
conversion of active  to sterile neutrinos through Shi-Fuller mechanism \cite{SF} in 
presence of leptonic asymmetry  has also been investigated \cite{RP1,RP2, RP3}. 
In this scenario, sterile neutrino parameters required to reproduce the X-ray line of $\sim$3.5 keV
are consistent with main cosmological parameters inferred from present cosmological 
measurements, Local Group and high-z galaxy count constraints and successfully solve the missing satellite 
and  too-big-to-fail problems \cite{Aba-RP,RP4,RP5}. 
Some tension with Ly-$\alpha$ data still exists (at 2.5 $\sigma$ level) \cite{Merle1}. 
This tension however, which could be related to some uncertainties in theoretical modelling of the intergalactic medium (IGM)
and the associated numerical methods \cite{Aba-RP,ly}, is not strong enough to rule out the RP scenario. \\
The keV sterile neutrino DM production by particle decays has been also extensively discussed \cite{PD1,PD2,PD3,PD4,WhitePaper}.
A particularly interesting case is the DM sterile neutrino production by scalar decay (SDP). 
This process involves a generic scalar singlet with the vacuum expectation value (vev) $<S>$ that could be 
produced via SM Higgs interactions. Depending on the strength of the Higgs coupling $\lambda_{H}$, the singlet scalar can be  produced 
like WIMPs  via {\it freeze-out} \cite{Petraki,Kusenko1,Kusenko2} or like  ``Feeble Interacting Massive Particles" (FIMPs) via {\it freeze-in} 
\cite{Merle1,Merle2} mechanisms
and must couple with the right-handed neutrino fields through Yukawa interaction, leading to sterile neutrino Majorana masses $m_{N}=y_{k} <S> $, where $y_{k}$ is the Yukawa coupling strength. Ref. \cite{Merle3} presents a complete treatment of the SDP mechanism for the whole parameter space,
giving the general solution on the level of momentum distribution function. 

Other proposed mechanisms are the production via interactions with the
inflaton field \cite{Shapo,Bez}, or production from pion decays \cite{pion}. \\
The coupled DE models (CDE) in which the DM particles, in addition to the gravitational interaction, 
have an interaction mediated by the DE scalar field have been also studied. 
A classification of these models can be found in Ref. \cite{grav2}. 
%They are known for providing a solution to the coincidence problem \cite{CDE1}.
The strength of coupling modifies the shape and amplitude of cosmological perturbations \cite{CDE2}, affecting the growth rate of cosmological structures \cite{CDE3}. Moreover, the strength of the coupling is degenerate with the amount of DM energy density, with impact on 
different cosmological parameters, including the Hubble expansion rate \cite{CDE4} and equation 
of state of DE \cite{CDE5}.

So far, the  keV sterile neutrino DM properties 
have been addressed by evaluating their impact on the co-moving free streaming horizon,
that  relates on the average velocity distribution. 
However, for  such models characterised by a highly non-thermal momentum distribution, 
the average momentum is subject of uncertainties, leading to a fail of free-streaming horizon 
in constraining the sterile neutrino parameters \cite{Merle1a}. 
The existing constraints are in general obtained in linear theory under the 
assumption that sterile neutrinos are all of the DM \cite{Merle2,Aba14,Murgia}. \\
The aim of this paper is to place constraints on RP 
and SDP mechanisms  through their impact on distance-redshift relations and the growth of structures.
We consider models where DM is a mixture of CDM and sterile neutrino
produced via RP and SDP mechanisms and analyse if this mixture 
can be compensated by changes in cosmological parameters. \\
We use the existing measurements of the CMB gravitational potential,
of the baryon acoustic oscillation (BAO) and 
of the weak gravitational lensing of galaxies to discriminate between different sterile neutrino DM production mechanism through the impact on 
the acoustic scales, the small scale  fluctuations and 
the low-redshift probes.\\
The paper is organised as follows: Section 2 summarise the RP and SDP Boltzmann formalisms 
calculations. Section 3 describes the model parameters and the  methods involved in the analysis. Section 4 
presents the datasets. Section 5 presents our results and examine the consistency
and cosmological implications of sterile neutrino DM production mechanisms. The conclusions are summarised in Section 6.

\section{Sterile neutrino DM production mechanisms}

In this section we present the sterile neutrino DM production calculations. We compute the evolution of phase space distributions in an homogeneous and isotropic Friedman-Robertson-Walker universe employing 
the Boltzmann equation:
\begin{eqnarray}
\label{Btz}
{\hat L}[f]={\cal C}[f]  \,, 
\end{eqnarray}
where  $f$ is the phase space distribution,  ${\cal C}$ is the collision term which encodes the details of a specific sterile neutrino DM production mechanism  and ${\hat L}$ is the Liouville operator:
\begin{eqnarray}
\label{Lv}
{\hat L}=\frac{\partial}{\partial t} - {\cal H}p\frac{\partial}{\partial p} \,,
\end{eqnarray}
where $p$ is the particle momentum and $\cal H$ is the Hubble function.
In order to bring Eq. (\ref{Lv}) into a more convenient form, we perform the following transformation of variables  \cite{Merle3}:
\ba
\label{var}
t &\rightarrow& r=r(t,p) \,, \\ \nonumber
p &\rightarrow&  \xi=\xi(t,p) \,.
\ea
Exploiting the correspondence between temperature $T$ and time $t$ and by using the conservation of the comoving entropy, 
the above transformations can be written in the form (for details see Appendix A.2 from Ref. \cite{Merle3}):
\ba
\label{new_var}
r &=& \frac{m_0}{T} \, , \nonumber \\
\xi &= &\left( \frac{g_s (T_0)}{g_s(T)}\right)^ {1/3} q \,,
\ea
where  $q=p/T$  is the co-moving momentum and $g_s(T)$ is the effective number of relativistic entropy degrees
of freedom. We choose $m_0=T_0=m_{h}$ where $m_{h}=125$ GeV is the Higgs boson mass. 
In terms of the variables given in  Eqs.(\ref{new_var}), the Liouville operator reads as:
\ba
\label{new_Lv}
{\hat L}= {\cal H} r \left( \frac{T g^{'}_s(T)}{3g_s(T)} +1 \right)^{-1} \frac{\partial}{\partial r} \,,
\ea
and the time-temperature relation is given by:
\ba
\label{t-T}
\frac{dT}{d t} = -{\cal H} T \left( \frac{T g^{'}_s(T)}{3g_s(T)} +1 \right)^{-1}\,,
\ea
where $ ^{'}$ denotes the derivative with respect to the temperature $T$. We used the fitting formulas from Ref.\cite {dgf} to compute the temperature evolution of the effective number of relativistic entropy degrees of freedom $g_s(T)$ and its derivative $g_s^{'}(T)$.

\subsection{Sterile neutrino resonant production (RP)}

The Boltzmann equation describing the sterile neutrino RP in terms of  variables given by Eqs. (\ref{new_var}) can be written as:
\ba
\label{Btz_RP}
{\cal H} r \left( \frac{T g^{'}_s(T)}{3g_s(T)} +1 \right)^{-1} \frac{\partial}{\partial r} f_{\nu_s}(r, \xi) \simeq 
\Gamma ( f_{\nu_{\alpha}} \rightarrow   f_{\nu_{s} })
\left[ f_{\nu_{\alpha}} (r, \xi) - f_{\nu_s}(r, \xi)\right ] \,.
\ea
There is similar equation for antineutrinos ${\bar \nu}_{\alpha}$.
In the above equation  $f_{\nu_{\alpha}}$ ($\alpha = e, \mu, \tau$) is the active neutrino momentum distribution function,
$f_{\nu_s}$ is the sterile neutrino momentum distribution function and $\Gamma(\nu_{\alpha} \rightarrow \nu_s)$
is the sterile neutrino effective production rate \cite{sn2,RP1}.
\begin{equation}
\Gamma(\nu_{\alpha} \rightarrow \nu_s)\approx
\frac{1}{4}\Gamma_{\nu_{\alpha}}(p,T) \sin^2 2\theta_M \,,
%\left < P_m(\nu_{\alpha} \rightarrow \nu_{s}) \right > \,,
\end{equation}
where   $\Gamma_{\nu_{\alpha}}$ is the collision rate and $\theta_M$  is the effective  matter mixing angle:
\begin{equation}
\label{sin2_m}
\sin^2 2\theta_M
 = \frac{\Delta^2(p) \sin^2 2\theta}
{\Delta^2(p) \sin^2 2\theta+ D^2(p)+
[ \Delta(p)\cos 2\theta+V^{L}-V^{T}] ^2} \,.
\end{equation}
Here $\theta$ is the vacuum mixing angle, $\Delta(p)=\delta m^2/2 p$ is the vacuum oscillation factor, 
$D(p)=\Gamma_{\nu_{\alpha}}(p)/2$ is the quantum damping rate, 
$V^T$ is the thermal potential and $V^L$ is the asymmetric lepton potential.
For temperatures characteristic to the post weak decoupling era (T $<$ 3 MeV),
the contribution of the thermal potential is very small and can be neglected.
In the presence of a primordial lepton asymmetry $V^L$ is given by:
\begin{eqnarray}
V^{L}=2 \sqrt{2}\zeta_R(3)\pi^{-2}G_FT^3{\cal L}_{\alpha} \,.
\end{eqnarray}
where $G_F$ is the Fermi constant, $\zeta_R(3)$ is the Reimann  function of 3 and ${\cal L_{\nu_{\alpha}}}$ is the potential lepton number corresponding to the active neutrino flavour $\alpha$:
\begin{flushleft}
\begin{eqnarray}
{\cal L_{\nu_{\alpha}}} \equiv 2 L_{\nu_{\alpha}} +\sum_{\beta \ne \alpha} L_{\nu_{\beta}} \,,
\hspace{0.3cm}
L_{\nu {\beta}} = \left( \frac{1}{12 \zeta_R(3)}\right) 
\left(  \frac {T_{\nu}}  {T_{\gamma}} \right)^3
[\pi^ 2 \zeta_{\nu_{\alpha}} + \zeta_{\nu_{\alpha} }] &
\hspace{0.5 cm} \beta=(e, \mu, \tau)
\end{eqnarray}
\end{flushleft} 
In the above equation 
$ \pm \zeta_{\nu_{\alpha}}$ is $\nu_{\alpha}/ {\bar \nu}_{\alpha}$ 
chemical potential, $L_{\nu_{\beta}}$ is the  lepton number 
and $T_{\nu}$ is the present temperature of the neutrino background $[T_{\nu} /T_{\gamma} = (4/11)^{1/3}]$.
The MSW condition for the resonant scaled neutrino momentum $\zeta_{res}$
is given by:
\begin{equation}
\label{xi_res}
\zeta_{res}=\left( \frac{g_s(T_0)}{g_s(T)}\right)^{1/3} \left( \frac{p}{T}\right)_{res} 
 \approx \left( \frac{\delta m^2 \cos 2\theta}
{4 \sqrt{2} \zeta_R(3)\pi^{-2}G_F {{\cal L_{\nu_{\alpha}}}}} \right) T^{-4}\,, 
\end{equation}
where $\delta m^2=m_2^2-m_1^2\approx m_2^2$ is the difference of the squares of  sterile neutrino and active neutrino mass eigenvalues, $T$ is the plasma temperature and $\theta$ is the vacuum mixing angle. 
 The evolution of the potential lepton number when the resonant 
active neutrino momentum sweeps from 0 to $\zeta_{res}$ is then given by:
\begin{eqnarray} 
\label{lpot}
{\cal L}_{\nu_{\alpha}}(\zeta_{res})={\cal L}^{init}-\frac{1}{2 
{\zeta_R(3)}}\left( \frac{T_{\nu}}{T_{\gamma}}\right)^3 
\int^{\zeta_{res}}_{0}  (1-e^{-\pi \gamma/2}) 
f_{\nu_{\alpha}}  {\rm d} \zeta \,,
\end{eqnarray}
where $f_{\nu_{\alpha}}$ is the initial neutrino
Fermi-Dirac momentum distribution and $\gamma$ is the dimensionless adiabaticity parameter \cite{LZ}.
The sterile neutrino number density $n_{\nu_{s}}(r)$ and the sterile neutrino physical energy density  $\Omega_{\nu_s}h^2$ are then given by:
\begin{equation}
\label{ndens}
n_{\nu_s}(r)=\frac{N}{2 \pi^2}\frac{g_s(T)}{g_s(T_0)} \left( \frac{m_0}{r} \right)^3 \int^{\xi}_{0} {\rm d} \xi \, \xi^2 f_{\nu_s}(\xi,r) \,,
\end{equation}
\begin{equation} 
\label{omnus}
\Omega_{\nu_s}h^2=\frac{s_0}{s(r)} \frac{m_{\nu_s} n_{\nu_s}(r)}{\rho_c/h^2} \,,
\end{equation}
where $s(r)$ is the co-moving entropy density, $s_0$=2891.2 cm$^{-3}$ is the entropy density at the present time 
and $\rho_c/h^2$=1.054 10$^{-2}$ MeV cm$^{-3}$ is the critical density in terms of reduced Hubble constant $h$.\\
The sterile neutrino RP mechanism is parameterised with respect to  the sterile neutrino mass 
$m_{\nu_s}$, the matter mixing angle $\sin^2 2\theta$, and the initial lepton asymmetry of each active neutrino flavour, ${\cal L}_{\nu_{\alpha}}$. 
We simultaneously evolve Eqs. (\ref{t-T}), (\ref{Btz_RP}), (\ref{xi_res}) and (\ref{lpot}) to obtain  the active and sterile 
neutrino phase-space distributions in the expanding universe for the entire range of resonant scaled neutrino momentum. The details of this computation can be found in Ref. \cite{Popa}.
\begin{figure}%
\centering
\subfigure{%
\includegraphics[width=7cm,height=7cm]{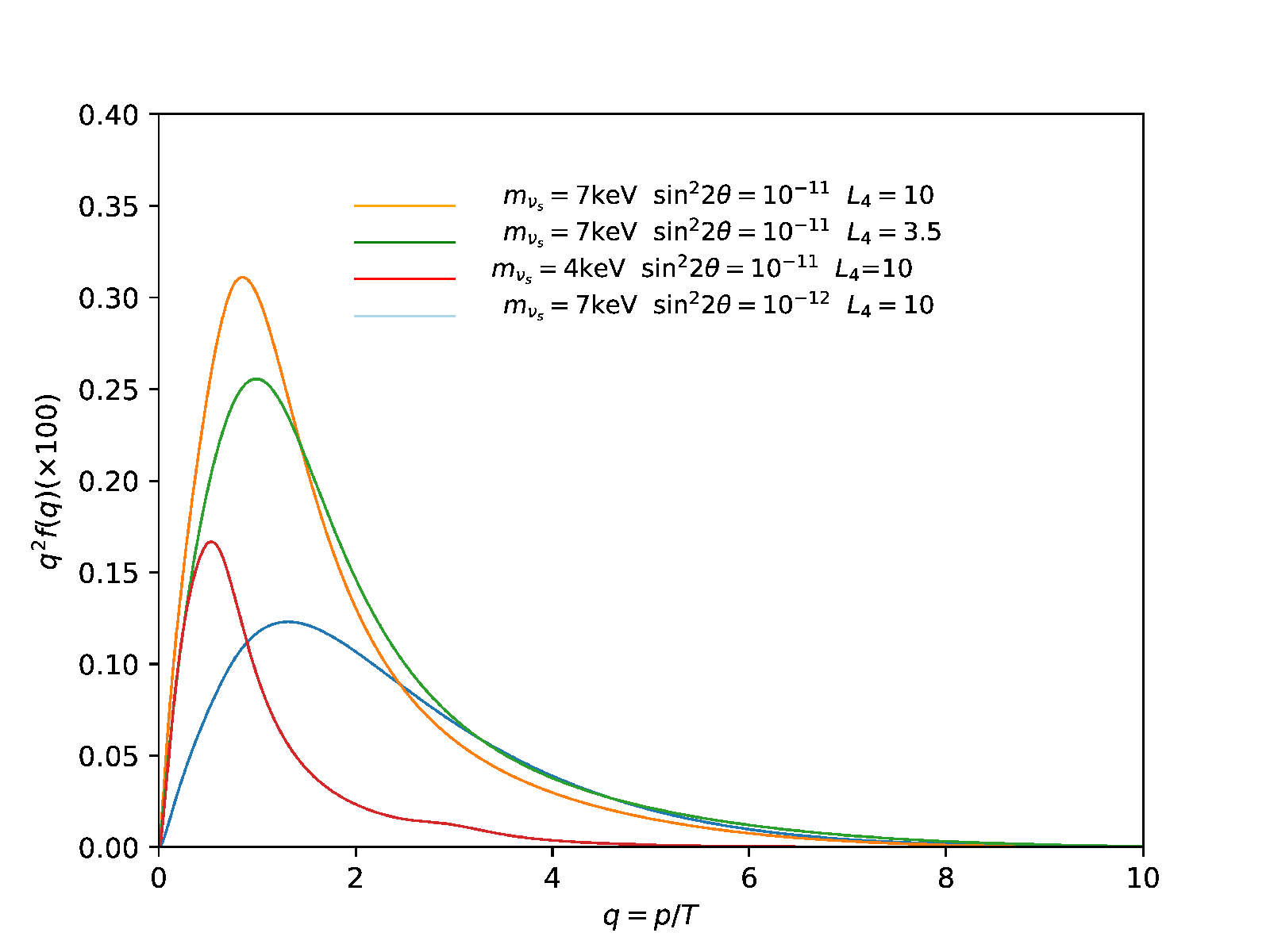}}%
\qquad
\subfigure{%
\includegraphics[width=7cm,height=7cm]{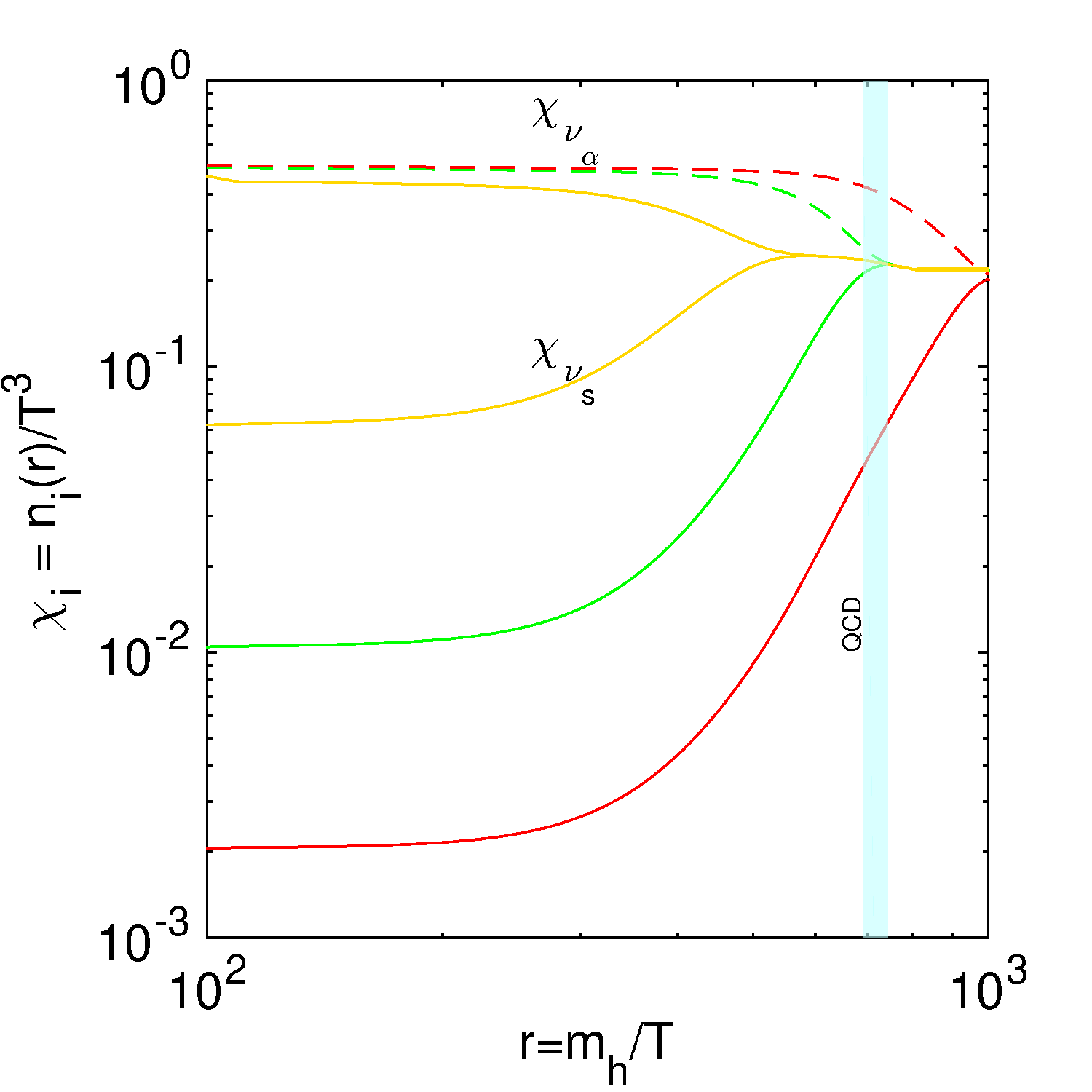}}%
\caption{{\it Left}: The dependence of the sterile neutrino final momentum distributions on the co-moving momentum 
$q=p/T$  for different values of $m_{\nu_s}$, $\sin^2 2\theta$ and $L_4 = 10^4 \times L_{\nu}$. \\
{\it Right}: The dependence of the abundances of sterile neutrino (continuous lines) and active neutrino (dashed lines)  on the time parameter $r=m_h/T$ 
for the same models presented in the left panel. The vertical light blue line indicates the temperature of the QCD phase transition ($T_{QCD} =173 \pm 8 {\rm MeV}$ \cite{Aba-RP}). 
Other parameters are fixed to: $\Omega_b h^2=0.0226$, 
$\Omega_ch^2=0.112$, $\Omega_{\nu}h^2=0.00064$, $H_0$=70 km s$^{-1}$Mpc$^{-1}$ and $\Omega_K=0$.}
\end{figure}
Fig.~1 presents in the left panel the dependence of the sterile neutrino final momentum distributions on the co-moving momentum $q=p/T$  for different values of $m_s$, $\sin^2 2\theta$ and ${\cal L}_{\nu_{\alpha}}$.  
The effect of increasing ${\cal L}_{\nu_{\alpha}}$ when $m_s$ and $\sin^2 2\theta$  are fixed is the increase of the averaged co-moving momentum, leading to a larger cutoff scales in the gravitational potential and matter power spectra. 
The same behaviour is present when  $\sin^2 2\theta$ and ${\cal L}_{\nu_{\alpha}}$ are fixed and $m_s$ is increased. 
A larger value of $\sin^2 2 \theta$ leads to larger sterile neutrino production rates. 
The resonance occurs in this case at a higher temperature and smaller averaged co-moving momentum. 
The same behaviours are shown by these models in the right panel from Fig.~1 that presents the evolution with time parameter $r=m_h/T$ of active and sterile neutrino abundances ${\chi }(r)=n(r) / T^3$, where $n(r)$ are corresponding number densities. 

There are few shortcomings related to this computation. The calculation  of sterile neutrino momentum distribution
is based on the semi-classical Boltzmann formalism which is accurate only if the collisions dominate the neutrino interactions. This restricts  the sterile neutrino parameter space to $\Delta(p) \sin^2 2 \theta / D(p)<1$ \cite{Vanu}, 
which we took into account in the cosmological analysis. 
Our computation  is also restricted  to the mixing 
of $\nu_s$ with one active flavour $\nu_{\mu}/{\bar \nu_{\mu}}$, 
ignoring the mixing with other flavours that may have similar momentum distributions \cite{mix_more}.
We also assume the same initial lepton asymmetry of each  $\nu/{\bar \nu}$  flavour and ignore the redistribution of the  lepton asymmetry during the QCD phase-transition and the opacities of active neutrinos. 
However, in Ref. \cite{Vanu} it is shown that these approximations do not significantly affect the sterile neutrino momentum distribution.\\
We used the sterile neutrino production code {\texttt sterile-dm}  
from \cite{Vanu}, that includes the initial lepton asymmetry redistribution and neutrino opacity correction, 
to test our production code. We find a good agreement between the momentum distributions.
We then  implement the momentum distributions obtained with {\texttt sterile-dm} code in our production code 
and find $\Omega_{\nu_s}h^2$ in agreement up to $\pm $5\%, for a large range of parameter space.
We also find that the  sterile neutrino momentum distributions obtained with our code are in agreement with 
the similar momentum distributions presented in Refs. \cite{Schneider,Aba14}.

\subsection{Sterile neutrino production by the scalar decay (SDP) }

In the case of SDP mechanism, the evolution of momentum distributions for scalar, $f_S$,  and sterile neutrino, 
$f_{\nu_s}$, are obtained by solving the coupled Boltzmann equations:
\ba
\label{Btz_SDP}
{\hat L}[f_{S}]= {\cal C}^{S} \,, \hspace{0.3cm}  {\hat L}[f_{\nu_s}] = {\cal C}^{\nu_s} \,,
\ea
where ${\hat L}$ is the Liouville operator given in Eq. (\ref{new_Lv}) and 
${\cal C}^S$ and ${\cal C}^{\nu_{s}}$ are the scalar and sterile neutrino collision terms 
encoding the effects of different processes that contribute to their production.  
In this work we take the leading processes contributing to ${\cal C}^S$ and ${\cal C}^{\nu_s}$:
\ba
\label{SD_col}
{\cal C}^S & = & {\cal C}^{S}_{hh \leftrightarrow SS} + {\cal C}_{S \rightarrow \nu_s \nu_s}  \, \nonumber \\
{\cal C}^{\nu_s} & = & {\cal C}^{\nu_s}_{S \rightarrow \nu_s \nu_s} \,
\ea
where: ${\cal C}^{S}_{hh \leftrightarrow SS}$  describes the depletion of scalars due to the annihilation into pairs of SM Higgs particles and the reverse process,  ${\cal C}_{S \rightarrow \nu_s \nu_s}$ describes the decay of scalars into  pairs of sterile neutrinos and
${\cal C}^{\nu_s}_{S \rightarrow \nu_s \nu_s}$ describes the creation of sterile neutrinos from the decays of scalars. 
A detailed discussion regarding the contributions of different processes to the collision terms  can be found in Refs. \cite{Merle2,Merle3}.\\
With these assumptions, the SDP scenario is parametrised with respect to the sterile neutrino mass $m_{\nu_s}$, the scalar mass $M_S$, the strength of the Higgs coupling $\lambda_{H}$ and of the Yukawa coupling $y_{k}$. 
We use the explicit forms of the collision terms  given in Ref. \cite{Merle3} and
simultaneously evolve Eqs. (\ref{t-T}) and  (\ref{Btz_SDP}) to obtain  the scalar and sterile 
neutrino  momentum distributions in the expanding universe. The sterile neutrino number density and the  corresponding physical energy density are then obtained by using  Eqs. (\ref{ndens}) and (\ref{omnus}).
\begin{figure}%
\centering
\subfigure{%
\includegraphics[width=7cm,height=7cm]{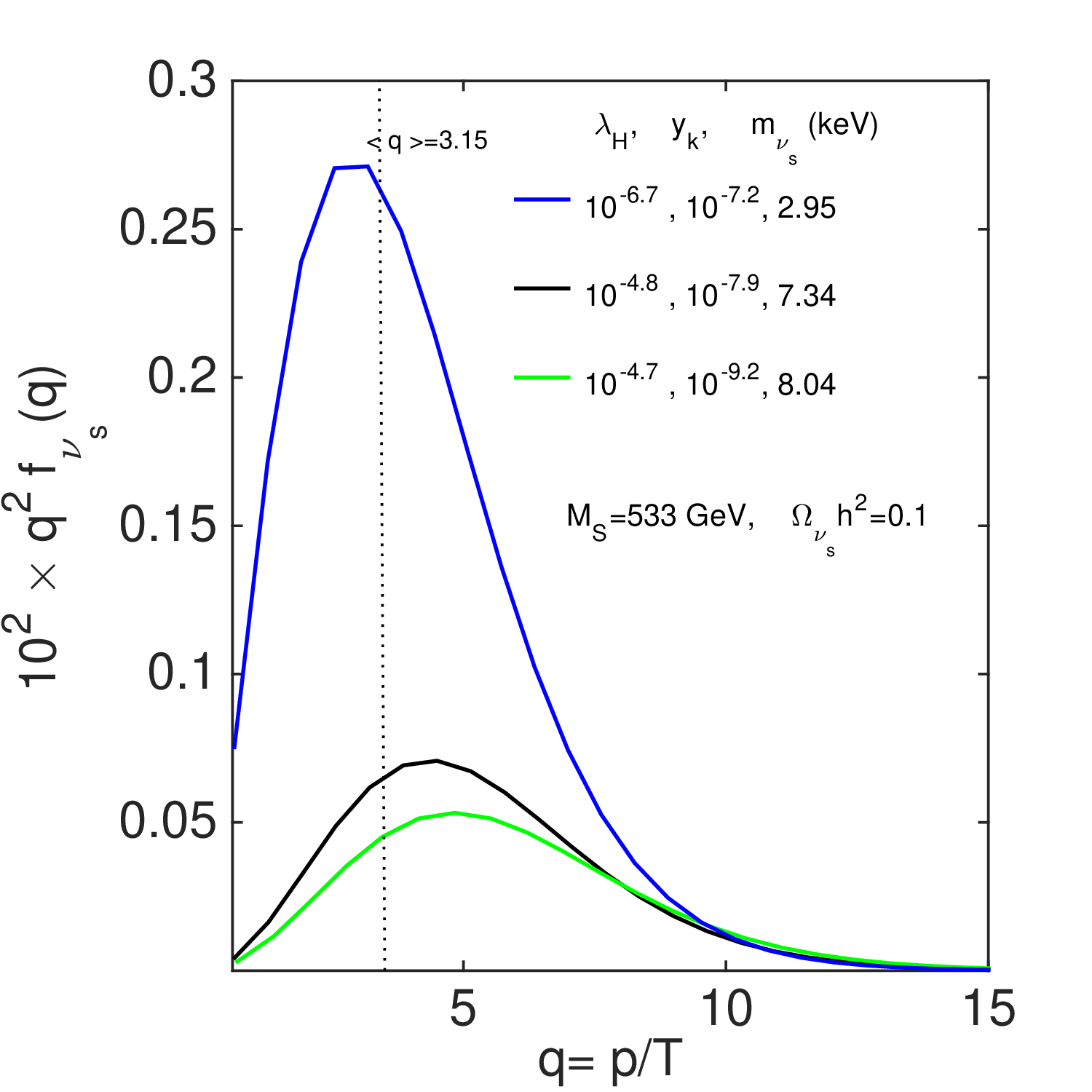}}%
\qquad
\subfigure{%
\includegraphics[width=7cm,height=7cm]{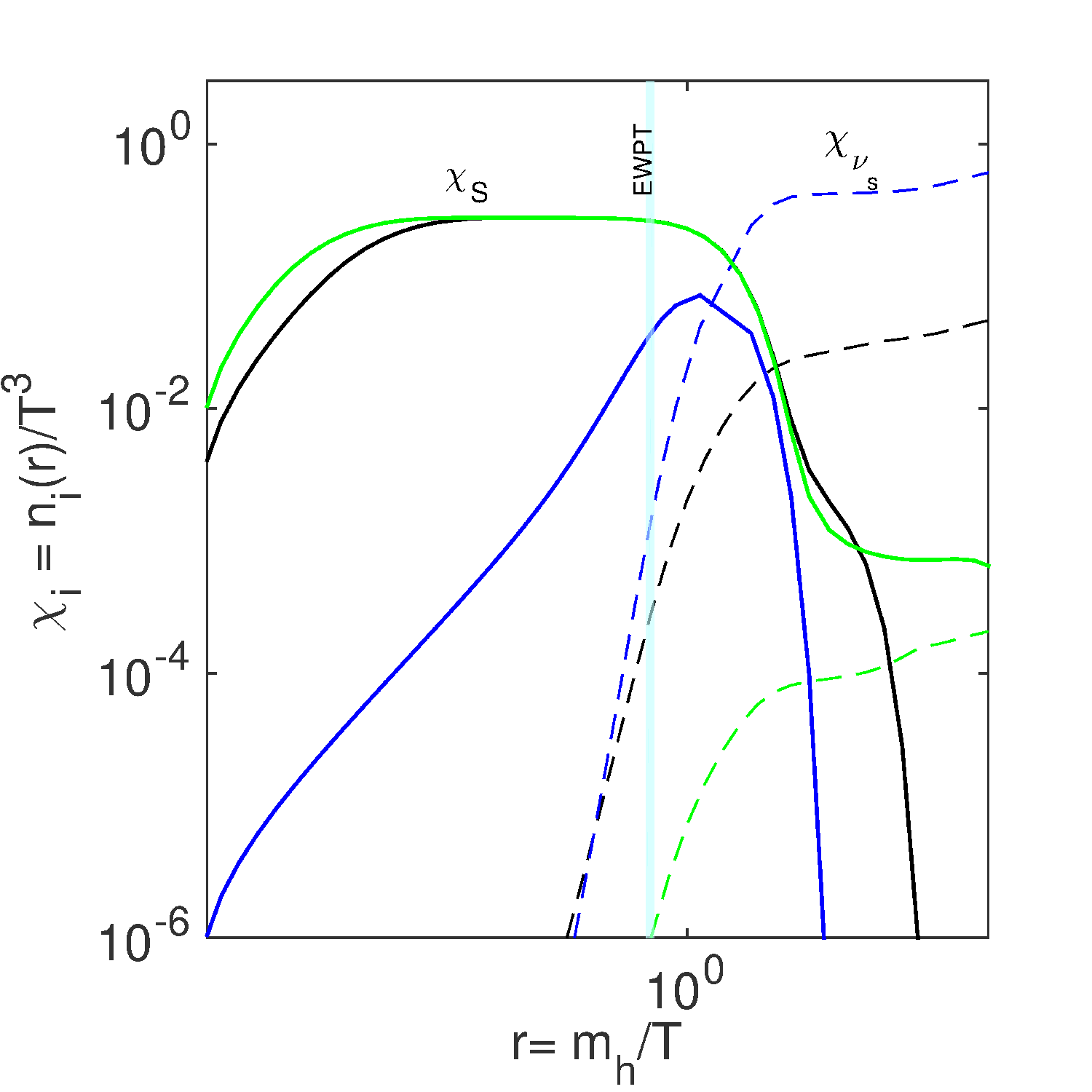}}%
\caption{{\it Left}:
The dependence of the sterile neutrino final momentum distributions on the co-moving momentum 
$q=p/T$  for scalar mass $M_S$=533 GeV (the best fit value) 
and different values of the Higgs coupling, $\lambda_H$, and  Yukawa coupling, $y_k$.
The dotted line indicates the value of the averaged co-moving momentum corresponding to sterile neutrino thermally produced.
{\it Right}: The dependence of the abundances of scalar (continuous lines) and sterile neutrino (dashed lines)  
on the time parameter $r=m_h/T$ for the same models presented in the left panel. The vertical blue line indicates the temperature of the electroweak phase transition. 
The distributions corresponding to the best fit parameters from our cosmological analysis 
are presented in both panels (black lines). Other parameters are fixed to: 
$\Omega_b h^2=0.0226$, 
$\Omega_ch^2=0.112$, $\Omega_{\nu}h^2=0.00064$, $H_0$=70 km s$^{-1}$Mpc$^{-1}$ and $\Omega_K=0$.  }
\end{figure}  
Left panel from Fig.~2 presents  
the dependence of the sterile neutrino final momentum distribution on the co-moving momentum $q=p/T$  for different values of $\lambda_{H}$ and $y_{k}$. 
The right panel shows the evolution with time parameter $r=m_h/T$ of scalar and sterile neutrino abundances ${\chi }(r)=n(r) / T^3$.  The distributions obtained for the best fit parameters are also presented by. 

We neglect in our computation the mixing between active and sterile neutrino and therefore the contribution of DW mechanism, shown to have a very small contribution to the sterile neutrino production \cite{Aurel1}.
We test the production code over a large parameter space and find 
that our distributions are in agreement with 
the similar distributions presented in Refs. \cite{Merle2,Merle3}.

\section{Parameterisation and methods}

{\it The baseline model} is an extension of the flat $\Lambda$CDM model, described 
by the following cosmological parameters: \\
$${\bf P}_{\rm \Lambda CDM}=\left\{ \Omega_bh^2 \,,\,\Omega_ch^2\,,\,\theta_s\,,\,\tau\,,\,
{\rm log}(10^{10} A_s)\,,\, n_s\,,\,\sum {m_{\nu}}\,,\,N_{eff}\right\} \,,$$
where: $\Omega_bh^2$ is the present baryon energy density, $\Omega_ch^2$
is present CDM energy density, $\theta_{MC}$ the ratio of sound horizon to angular diameter distance at decoupling, $\tau$ is the optical depth at reionization, $A_s$ and $n_s$ are amplitude and spectral index of primordial power spectrum of curvature  perturbations at pivot scale $k=0.05$ Mpc$^{-1}$, $\sum{m_{\nu}}$ is 
the total mass of three active neutrino flavours 
and $N_{eff}$ the number of relativistic degrees of freedom that parametrise
the contributions from any non-interacting relativistic particles. 
In the SM with three active neutrino flavours, $N_{eff}=$3.046 due to non-instantaneous 
decoupling corrections \cite{mangano}.  \\
{\it The RP mechanism model} includes, in addition to the baseline model parameters, the following parameters:
$${\bf P}_{\rm RP}=\left\{ \zeta_{\nu}\,,\,m_{\nu_s}\,,\, \sin^22\theta\right\}\,,$$ \\
where: $\pm \zeta_{\nu}$ is the total chemical potential of three degenerated active $\nu/{\bar \nu}$ species, $m_{\nu_s}$ is the sterile neutrino mass and
$\sin^2 2\theta$ is the mixing angle.\\
{\it The SDP mechanism model} extend the baseline model by including the following parameters:\\
$$ {\bf P}_{{\rm SDP}}=\left\{ m_{\nu_s}\,,\,M_{S}\,,\,y_{k}\,,\, \lambda_H \right\} \,,$$ 
where: $m_{\nu_s}$ is the sterile neutrino mass, $M_S$ is the scalar mass, $y_k$ is the Yukawa strengths coupling and 
$\lambda_H$ the  Higgs strengths coupling.\\
The sterile neutrino mass fraction is a derived parameter, $f_S=\Omega_{\nu_s}/\Omega_c$,
where  the sterile neutrino energy density $\Omega_{\nu_{s}} $ is computed by using Eq. (\ref{omnus}).
The matter energy density in RP and SDP scenarios is then given by 
$\Omega_{m}=\Omega_c+\Omega_b+\Omega_{\nu}+\Omega_{\nu_s}$. 

We modify the baseline Boltzmann code \texttt{camb}\footnote{http://camb.info} \cite{camb} to allow 
the calculation of sterile neutrino DM production formalisms presented in the previous section.\\
{\it Non-linear corrections:} 
We use the \texttt {halofit} model \cite{Smith,Taka} implemented in the  \texttt {camb} code
to account for the  non-linear effects in CMB anisotropy and lensing potential power spectra.\\
{\it Recombination:} The process of recombination determines the size of the sound horizon at this epoch, 
affecting the characteristic angular size of the CMB fluctuations and the diffusion dumping scale.
 We use the recombination model developed in Ref. \cite{recfast} and further improved for full numerical 
implementation in the \texttt {recfast}\footnote{http://www.astro.ubc.ca/people/scott/recfast.html} 
code \cite{recfast_up}.\\
{\it Nucleosynthesis:} The model of the Big Bang Nucleosynthesis (BBN)  gives 
the relation between helium mass fraction, $Y_P$, photon-to-baryon ratio, $\rho_{\gamma}/ \rho_b$, 
and the number of relativistic degrees of freedom, $N_{eff}$.
In the case of RP mechanism, the leptonic asymmetry 
increases the radiation energy density parametrised by
variation of the number of relativistic degrees of freedom $\Delta N_{eff}$:
\begin{equation}
\label{delta_neff}
\Delta N_{eff} (\zeta_{\nu})= 3 \left[\frac{30}{7} \left(\frac{\zeta_{\nu}}{\pi}\right)^2 
+\frac{15}{7} \left(\frac{\zeta_{\nu}}{\pi}\right)^4 \right] \,.
\end{equation}
The leptonic asymmetry also shifts the beta equilibrium between protons and neutrons 
with effects on $Y_P$ that decreases with the increase of $\zeta_{\nu}$.
The electron neutrino/antineutrino, $\nu_e/ {\bar \nu}_e$, phase-space distributions 
determine the rates of the neutron and proton interaction at BBN. 
In the RP model, the non-thermal $\nu_e/ {\bar \nu}_e$ spectra change these rates and hence the
$Y_P$ value over the case with thermal Fermi-Dirac spectrum \cite{aba04,Smith06}.
We use the PArthENoPE BBN code \cite{Pisanti} to set the value of $Y_P$.
For SDP model we compute the dependence of $Y_P$ on $\Omega_b h^2$ and $N_{eff}$. 
For the RP model we consider in addition the  effects on $Y_P$ and 
 $\zeta_{\nu}$ for the change of neutron and proton interaction rates.
 
 The parameter extraction from cosmological datasets
is based on Monte-Carlo Markov Chain (MCMC) technique.  
We modify the latest publicly available version of the package 
\texttt{CosmoMC}\footnote{\url{http://cosmologist.info/cosmomc/}} 
\cite{cosmomc} to sample from the space of cosmological and sterile neutrino production 
mechanism parameters and generate estimates of their posterior mean and  confidence intervals. \\
%We assume a flat Universe and uniform priors for all parameters adopted in the analysis in intervals listed in Tab.~1 and %Tab.~2.  The Hubble expansion rate $H_0$ and sterile neutrino energy density  $\Omega_{\nu_s}h^2$
%are derived parameter in our analysis. We constrain their values  to reject the extreme/unlikely models. \\
We first run the modified versions of {\texttt CosmoMC} and \texttt {camb} 
setting  to  zero the additional parameters for RP and SDP models.
In both cases we find good consistency between our bounds and the existing constraints 
for $\Lambda$CDM model \cite{Planck16}. 
We use the default convergence settings implemented in {\texttt CosmoMC}:
${\rm MPI \_Limit \_Converge = 0.025 }$ and ${ \rm MPI\_Limit\_Converge\_Err = 0.2}$.
With these choices the {\texttt CosmoMC} run stops when  the confidence interval 
for each parameter at 95\% C.L. is accurate at $0.2\,\sigma$.  
This error can be reduced, but in this case the computing time increases to reach the convergence limit.
This become critical for non-standard models, as RP and SDP, for which the execution time is larger 
than in the standard case because of numerical evolution of momentum distributions in \texttt {camb}. \\
We use the same convergence criteria and made few test runs for RP and SDP models to optimise the prior intervals and 
sampling. 
The final runs are based on 120 independent channels for each model, reaching the convergence criterion 
$(R-1) \simeq 0.01$ for RP model and $(R-1)=0.007$ for SDP model. The $(R-1)$ criterion is defined as 
the ratio between the variance of the means and the mean of variances for the second half of chains \cite{cosmomc}. 

We assume a flat Universe and uniform priors for all parameters adopted in the analysis in the 
intervals listed in Tab.~1 and Tab.~2.  
The Hubble expansion rate $H_0$ and sterile neutrino energy density  $\Omega_{\nu_s}h^2$
are derived parameters in our analysis. We constrained $H_0$ values  to reject the extreme models 
and restrict the values of $\Omega_{\nu_s}h^2$ to the  interval of $\Omega_ch^2$.
The sterile neutrino mass lower limit  is restricted by the  Tremain-Gunn bound \cite{Tremain} while 
the upper limit is restricted by the non-detection of emission lines from X-ray observations \cite{Xray}.\\

\begin{table}
\caption{ Priors and constraints for the $\Lambda$CDM-ext parameters adopted in the analysis. 
All priors are uniform in the listed intervals. We assume a flat Universe.}
\begin{center}
\begin{tabular}{|l|c|}
\hline 
                 Parameter&Prior  \\
\hline
$\Omega_bh^2$&     [0.005,\,0.1]       \\
$\Omega_ch^2$&     [0.001,\,0.5 ]       \\
$100\theta_s$ & [0.5,\,10] \\
$\tau$& [0.01,\,0.9] \\
${\rm log}(10^10 A_s)$ & [2.5,\, 5]\\
$n_s$& [0.5,\,1.5]\\
$m_{\nu}(\rm eV)$ &[0,\,6]\\
$N_{eff}$ &[3.046,\,8]\\
\hline
$H_0({\rm km\,s}^{-1}{\rm Mpc}^{-1})$& [20,\,100] \\
\hline
\end{tabular}
\end{center}
\end{table}

\begin{table}
\caption{ Priors and constraints on the additional parameters for RP and SDP models. 
All priors are uniform in the listed intervals.}
\begin{center}
\begin{tabular}{|l|c||l |c|}
\hline
RP Parameter & Prior& SDP Parameter& Prior\\ 
\hline
$m_{\nu_s}({\rm keV})$&[2,\,30]   & $m_{\nu_s}({\rm keV})$&[2,\,30]\\
10$^{10}\times \sin^2 2\theta$& [0.1\,,\,100] &$y_k$& [$10^{-10}\,,\,10^{-8}]$    \\  
$\zeta_{\nu}$&[-0.1,\,0.1]    & $\lambda_H$& [$10^{-8}\,,10^{-4}$]        \\
                     &                     & $M_S({\rm GeV})$& [10\,,\,1000]        \\
\hline
$\Omega_{\nu_s}h^2$&     [0.001\,,\,0.5]& $\Omega_{\nu_s}h^2$&[0.001\,,\,0.5]     \\          
\hline
\end{tabular}
\end{center}
\end{table}
%====================================================================
\section{Cosmological data}

For our cosmological analysis we use the following datasets:

{\it The CMB measurements:} We use the CMB angular power spectra from {\sc Planck} {\texttt 2015} release \cite{Pl15} 
and the {\sc Planck} likelihood codes \cite{PlanckXV} corresponding to different multipole ranges:
\texttt{Commander} for $2 \leq l\leq 29$,
\texttt{CamSpec}  $50\leq l\leq 2500$,
\texttt{LowLike} for $2\leq l\leq 32$ for polarization data
and \texttt{Lensing} for $40\leq l\leq 400$ of lensing data. 
As sterile neutrino production is expected to affect the redshift - distance relations and the growth of structures,
we include in the analysis 
the {\sc Planck} power spectrum of the reconstructed lensing potential \cite{Pl-lens}. 
We will refer tot the combination of these measurements as {\sc Planck}+lens dataset.

{\it Baryonic acoustic oscillations} (BAO): BAO measurements are low-redshift probes 
insensitive to non-linear effects because  their characteristic acoustic scale, of around 147 Mpc, is much larger than the scale of the 
virialized cosmological structures. Moreover, as BAO features in the matter power spectrum 
can be observed as a function of both angular and redshift separations \cite{Anderson14,Marin} 
these measurements are robust geometrical tests. We include in analysis the BAO characteristic parameter 
measurements from  Baryon Oscillation Spectroscopic Survey (BOSS)
LOWZ at $z_{eff}=$0.32 and CMASS at $z_{eff}$=0.57\cite{Marin}, BOSS DR12 at $z_{eff}=$~0.38, 0.51 and 0.61 \cite{Alam17}
and 6dF Galaxy Survey (6dFGS) at $z_{eff}=0.1$ \cite{6dF}.\\
We will refer to the combination of these measurements as BAO dataset.

{\it Cosmic shear}: Weak lensing of galaxies, or cosmic shear,  constraints the gravitational 
potential at redshifts lower than the CMB lensing. 
Presently, the cosmic shear measurements are available from several surveys 
\cite{Heymans, Jee,Hildebrandt}. We use the {\sc CosmoMC} implementation of  
one-year Dark Energy Survey (DES) \cite{DESc} 
cosmic shear measurements described in Ref. \cite{Troxel},  referred hereafter as DES dataset.
%====================================================================
\section{Analysis and results}

\subsection{Sensitivity of cosmological data to sterile neutrino mass and mass fraction}
%$m_{\nu_s}$ and $f_{S}$}

\begin{figure}
\centering
\subfigure{%
\includegraphics[width=7cm,height=7cm]{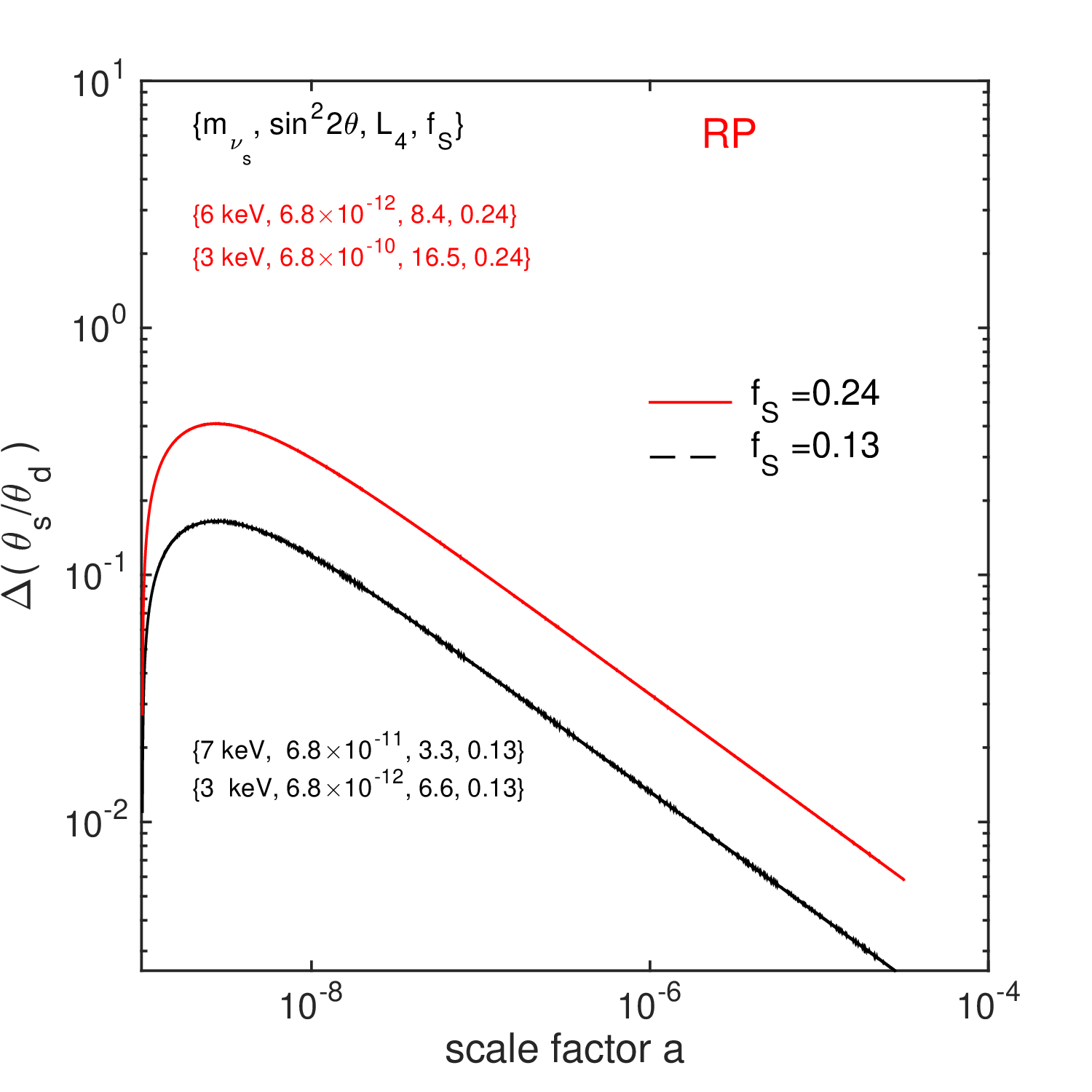}}%
\qquad
\subfigure{%
\includegraphics[width=7cm,height=7cm]{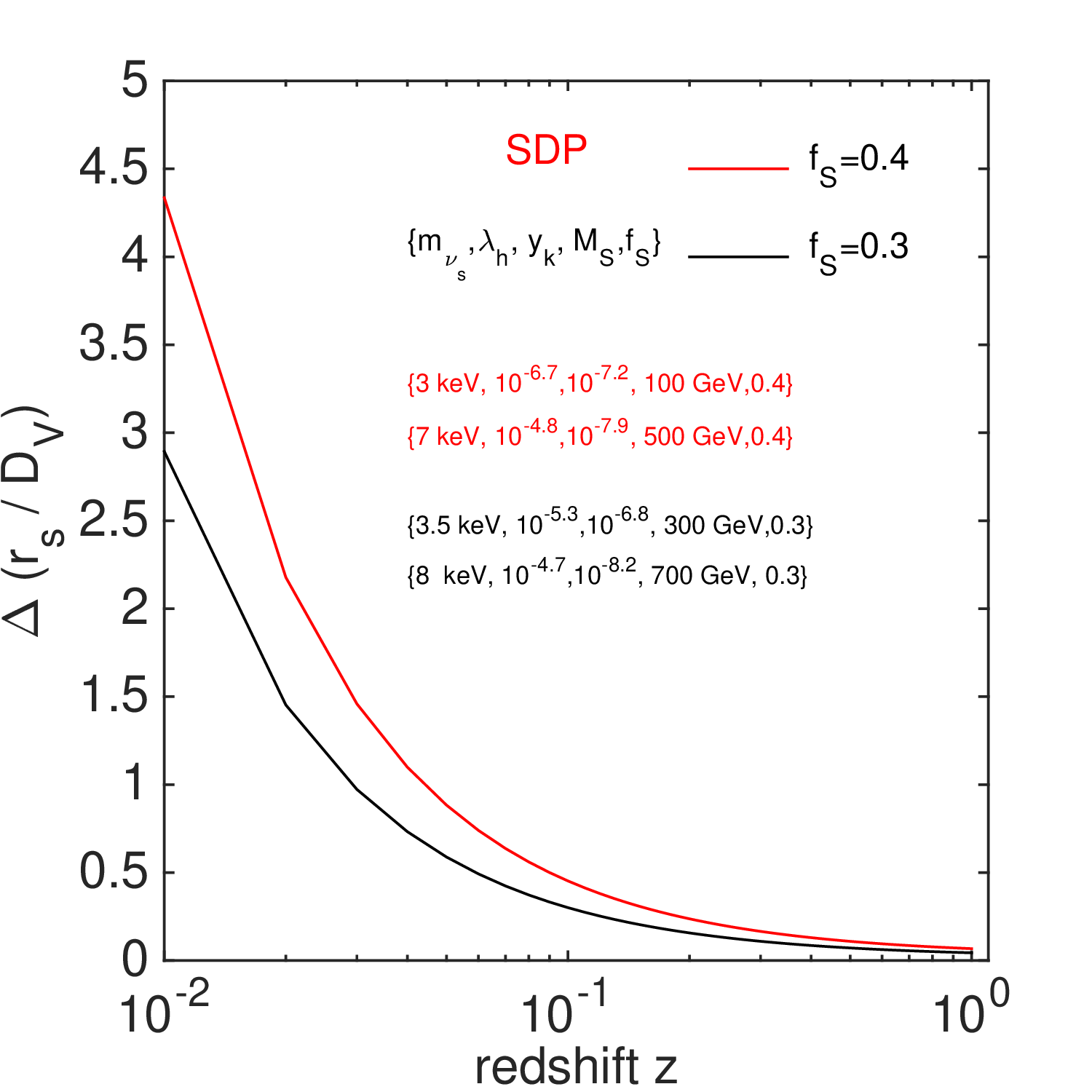}}
\caption{{\it Left}: The evolution to photon decoupling  of the variation 
$\Delta(\theta_s/\theta_d)$ for models sharing the same sterile neutrino mass fraction $f_{S }$ obtained in  the RP scenario. 
{\it Right}: The redshift dependence of the variation of BAO characteristic parameter, $\Delta (r_s(z_d)/D_V(z))$,
for models sharing the same sterile neutrino mass fraction $f_{S }$ obtained in  SDP scenario. 
The production mechanism  parameters are indicated for each case.  Other parameters are fixed to: 
$\Omega_b h^2=0.0226$, 
$\Omega_ch^2=0.112$, $\Omega_{\nu}h^2=0.00064$, $H_0$=70 km s$^{-1}$Mpc$^{-1}$ and $\Omega_K=0$.}
\end{figure}
A change in sterile neutrino mass fraction $f_S$ leads first to  a change in the  
Hubble expansion rate $H$.
At  the CMB photons decoupling ($T_{\rm cmb}$=0.26 eV) this change 
the sound horizon distance $r_s$ (that scales as $ 1/H$)
and in the photon diffusion dumping distance $r_d$ (that scales as $ \sqrt{1/H}$). 
The CMB anisotropy spectrum is sensitive to both $r_s$ and $r_d$ changes as projected angles over the  
co-moving angular distance, $D_A$, to the last scattering: $\theta_s=r_s/D_A$ and $\theta_d=r_d/D_A$. 
While a change in $\theta_s$ shifts the position of the acoustic Doppler peaks through the CMB anisotropy spectrum, 
a change in $\theta_d$ relative to $\theta_s$ modify its amplitude.
CMB can measure the expansion rate by measuring the ratio $\theta_d/\theta_s \sim \sqrt{H}$. \\
For models sharing the same value of $f_S$,  the ratio $\theta_d/\theta_s$  is a measure of 
the relativistic energy density at $T_{\rm cmb}$, usually  parametrised by $N_{eff}(T_{\rm cmb})$ \cite{Hou}.
The contribution of sterile neutrinos to $N_{eff}(T_{\rm cmb})$ encodes information on  their mass 
and production mechanism parameters that  lies in the momentum distribution function and 
can be computed by comparing the
sterile neutrino kinetic energy density to the energy density of  other 
relativistic particles in equilibrium at $T_{\rm cmb}$ \cite{Merle2}.
This contribution is very small in the case of sterile neutrinos produced via SDP mechanism 
since in this case sterile neutrinos 
cooled down at $T_{\rm cmb}$  and have smaller co-moving momenta when comparing to the RP mechanism.\\
Left panel from Fig.~3 presents the evolution to photon decoupling of the variation
$\Delta (\theta_s/\theta_d)$ obtained in RP case for models sharing  the same sterile neutrino mass fraction. 
The figure shows  that an accurate determination of $\theta_s/\theta_d$  breaks the degeneracy 
of Hubble parameter at $T_{\rm cmb}$,
leading to constrains on sterile neutrino mass and production parameters.

The BAO measurements lead to joint constraints on the co-moving angular diameter distance, $D_A(z)$,
and the Hubble parameter, $H(z)$, at smaller redshifts then CMB.  
At these redshifts, for models sharing the same sterile neutrino mass fraction, both $D_A(z)$ and $H(z)$ 
are degenerated.
The BAO observations typically constrain the quantity 
$r_s(z_{d})/{D_V(z)}$, where $r_s(z_{d})$ is the sound horizon distance at the drag epoch redshift $z_{d}$, 
when baryons and photons decouple, and $D_V(z)$ is given by:
\begin{eqnarray} 
\label{BAO-Dv}
D_V(z)=\left[ (1+z)^2 D^2_A(z)  \frac{cz} {H(z)} \right]^{1/3} \,,
\end{eqnarray}
where $c$  is the speed of light.  
The right panel from Fig.~3 presents the redshift dependence of 
the variation $\Delta (r_s(z_d)/D_V(z))$ for models sharing the same $f_{S }$ 
obtained in the SDP scenario. The figure shows that the BAO measurements break the degeneracy 
between these models, leading to constraints on sterile neutrino mass and production parameters.
%---------------------------------------------------------------------------------
\begin{figure}
\centering
\subfigure{%
\includegraphics[width=7cm,height=7cm]{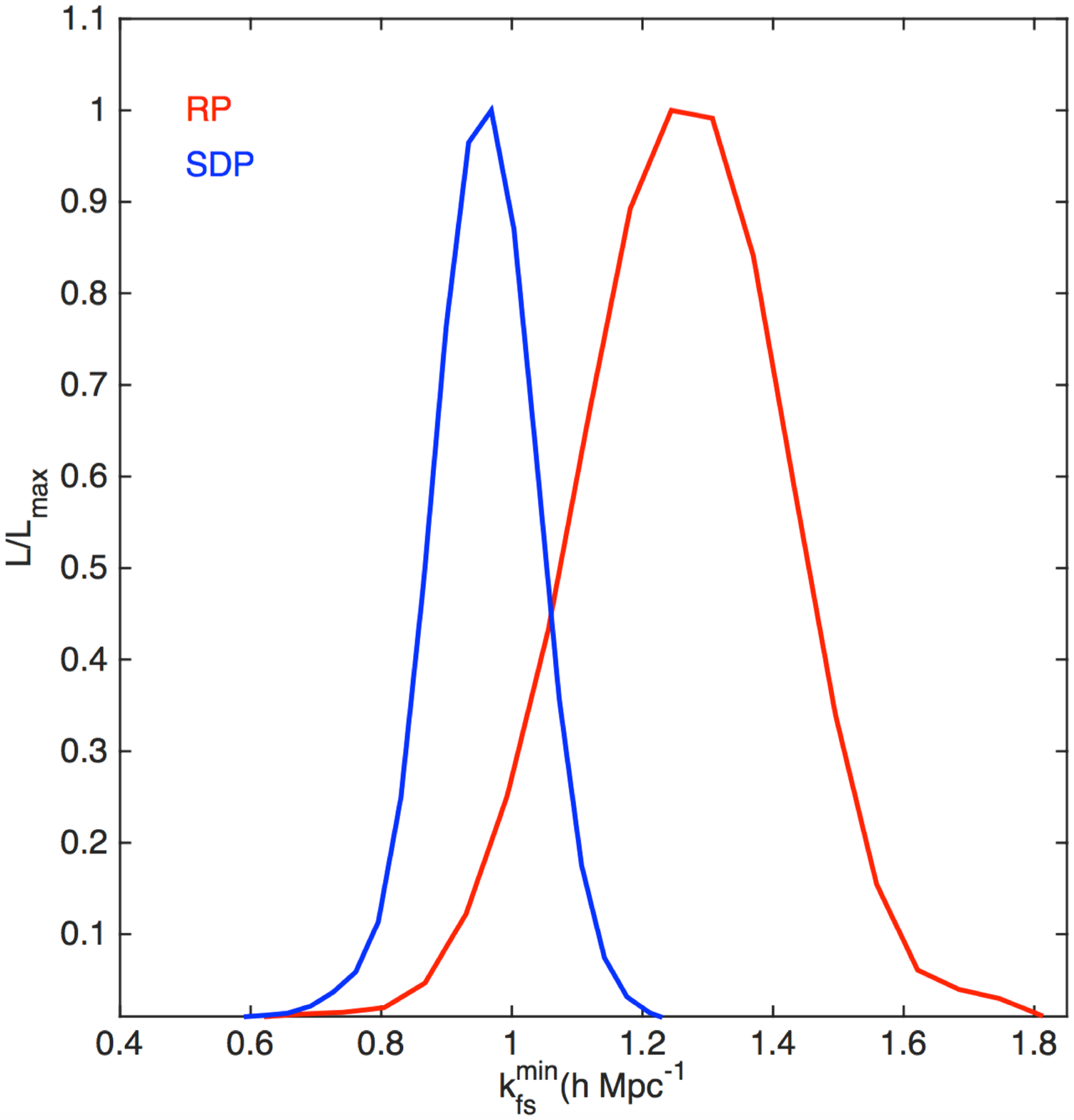}}
\qquad
\subfigure{%
\includegraphics[width=7cm,height=7cm]{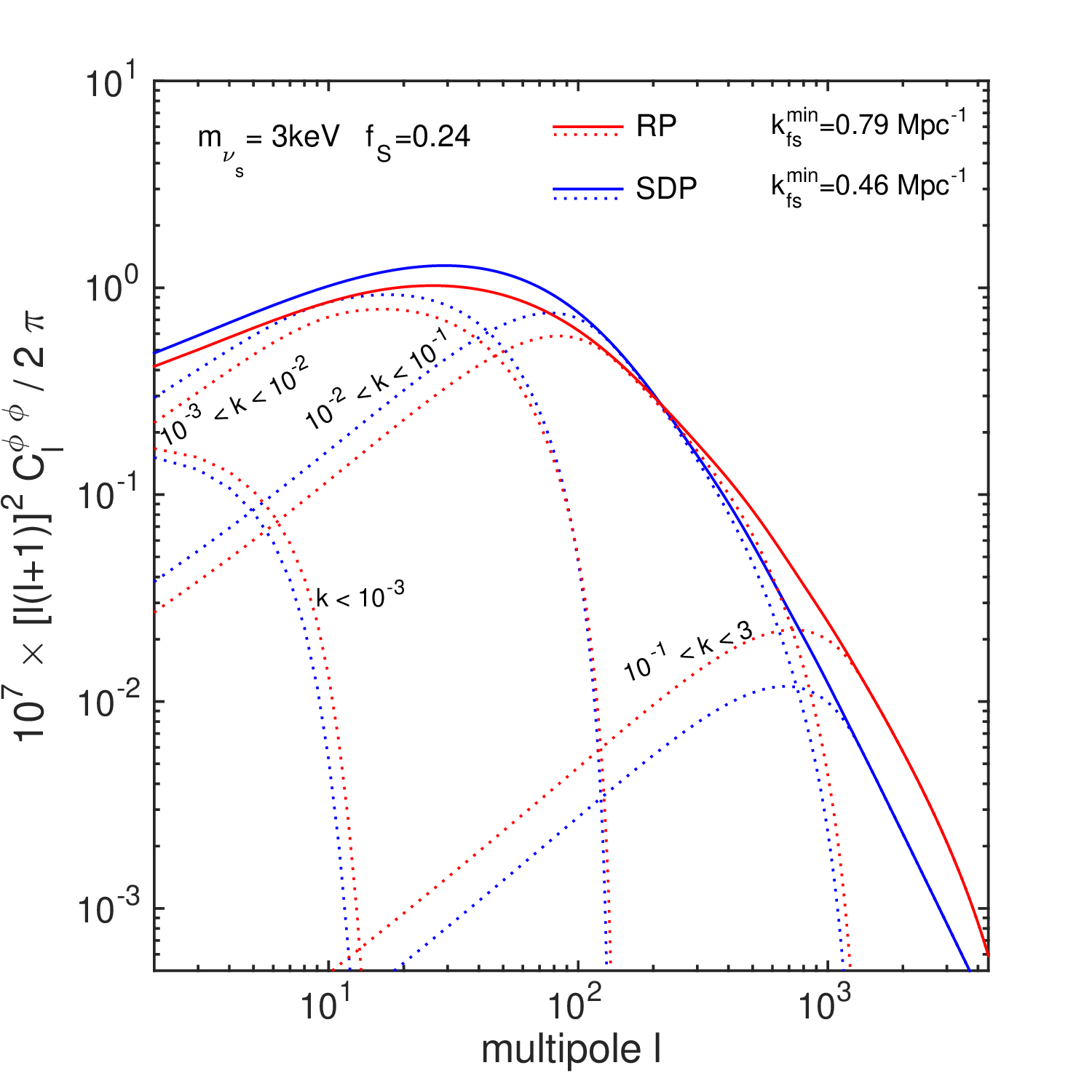}}
\caption{{\it Left}: The likelihood probability distributions of the estimated free-streaming 
horizon wave-number,
$k^{min}_{fs}$, from the fit of RP and SDP models with {\sc Planck}+lens+BAO+DES datasets.
{\it Right}: The deflection angle power spectra for 
models sharing the same values of $f_{S}$ and $m_{\nu_s}$ obtained in  RP and SDP scenarios. The contributions of different wave numbers $k$ (in Mpc$^{-1}$) to the deflection angle power spectra are also presented (dotted lines). 
RP and SDP mechanism parameters are indicated in Fig.~3. Other parameters are fixed to: 
$\Omega_b h^2=0.0226$, 
$\Omega_ch^2=0.112$, $\Omega_{\nu}h^2=0.00064$, $H_0$=70 km s$^{-1}$Mpc$^{-1}$ and $\Omega_K=0$.}
\end{figure}
%-----------------------------------------------------------------------------

The gravitational lensing of the CMB photons provides direct measurements 
on scales where the effects of sterile neutrino are expected to manifest. The largest scale affected 
is the present value of  co-moving free-streaming horizon given by \cite{Les3}:
\begin{equation}
\label{lambda}
\lambda^0_{fsh}= \int^{T_{prod}}_{T_{0}} \frac{< v(T) > }{ a(T)}  \frac{ {\rm d} t}{ {\rm d} T} {\rm d}T\,,
\end{equation}
where $<v(T)>$ is the velocity dispersion of sterile neutrinos, 
$T_{prod}$ is the sterile neutrino production temperature and $T_0$ is the present temperature.
We compute $\lambda^0_{fsh}$ for RP and SDP models  following the analytical 
approach  from Ref. \cite{Les3}: 
\begin{equation}
\label{lfsh}
\lambda^0_{fsh}=\frac{a_{nr}}{\sqrt{\Omega_R}H_0}\left(1+6 {\rm Arcsinh } \sqrt{ \frac{a_{eq}}{a_{nr}}}\right)\,,
\end{equation}
where $a_{eq}=\Omega_{R}/\Omega_m$ is the scale factor at matter-radiation equality, $a_{nr}=T_0/m_{\nu_s}$ 
is the scale factor at the time of sterile neutrino non-relativistic transition, $\Omega_R$ the is the radiation 
energy density and $\Omega_m$ is the matter energy density.
The analytical approach (\ref{lfsh}) assumes that Universe is completely radiation dominated until $a_{eq}$ and 
neglects the small contribution to the integral (\ref{lambda}) of the dark energy. 
We account for the entropy dilution from $T_{prod}$ until $T_0$ rescaling
$\lambda^0_{fsh}$ by a factor  $\xi_s^{1/3}=g_s(T_{prod})/g_s(T_0)$ \cite{Merle1}, where $g_s$
is the effective number relativistic entropy degrees of freedom ($g_s(T_{\rm QCD}) \approx 38.1$ for RP,
 $g_s(T_{\rm EWFT}) \approx 109.5$ for SDP and  $g_s(T_0) \approx 3.36$) and take 
into account the increase of $\Omega_R$ in RP model according to  Eq.(\ref{delta_neff}). \\
Left panel from Fig.~4 presents the likelihood probability distributions of the 
free-streaming horizon wave-number $k^{min}_{fs}$, obtained for our models.  
One should note that wave-numbers $k \sim 1$ hMpc$^{-1}$ correspond
to the typical size of dwarf galaxies \cite{Merle2},  while 
the  observations of  Ly-$\alpha$ absorption in spectra of distant quasars
are tracers of linear density fluctuations on scales  $0.1< k < 3$ hMpc$^{-1}$ \cite{aba2006}.

On the other hand, the power spectrum of the CMB projected gravitational potential,
$C^{\phi \phi}_l$, is sensitive to both geometry and growth of structures at wave-numbers $k>k^{min}_{fs}$. 
In the Limber approximation, $C^{\phi \phi}_l$ can be written as:
\begin{eqnarray}
\label{Clphi}
C^{\phi \phi}_l = \frac{8 \pi}{l^3}\int^{\chi^*}_0 {\rm d} \chi \, D_A(\chi)
\left( \frac{D_A(\chi^*)-D_A(\chi)}{D_A(\chi^*)D_A(\chi)} \right)^2\,P_{\Psi}(z(\chi) ,k=l/D_A(\chi)) \,,
\end{eqnarray}
where: $\chi$  is the co-moving coordinate distance, $\chi^*$ is the co-moving coordinate distance to the last scattering
surface, $k$ is the wave-number, $D_A(\chi)$ is the co-moving angular diameter distance
and  $P_{\Psi}(z,k)$ is the power spectrum of the gravitational potential. $P_{\Psi}(z,k)$ can be related 
to the power spectrum of matter density fluctuations, $P_{m}(z,k)$, through  the Poisson equation, leading to \cite{LwCh}:
\begin{eqnarray}
\label{Pm}
P_{\Psi}(z,k)=\frac{9 \Omega^2_{m} (\chi) H^4(\chi) }{8 \pi^2 } \frac{P_{m}(z,k)}{k}\,,
\end{eqnarray}
where $k$ is in units of Mpc$^{-1}$ and $P_{m}(k,z)$ is in units of Mpc$^{3}$. 
The deflection angle power spectrum of the CMB  lensing potential, as reported from the {\sc Planck} CMB lensing analysis
\cite{Lens}, is then given by $l^2(l+1)^2 C^{\phi \phi}_l$.\\
The right panel from Fig.~4  presents the deflection angle power spectra obtained for 
models sharing the same values of $f_{S}$ and $m_{\nu_s}$ obtained in  RP and SDP scenarios.
We indicate the contributions of different wave-numbers $k$ (in Mpc$^{-1}$) to the deflection angle power spectra. \\
The figure shows that for multipole range 
involved in this analysis, $40\leq l\leq 400$,  the deflection angle power spectrum of the CMB  lensing potential
is sensitive to both sterile neutrino production mechanisms,
with an increased value of the wave-number of power suppression in the RP case.
 \begin{figure}%
\centering
\subfigure{%
\includegraphics[width=7cm,height=7cm]{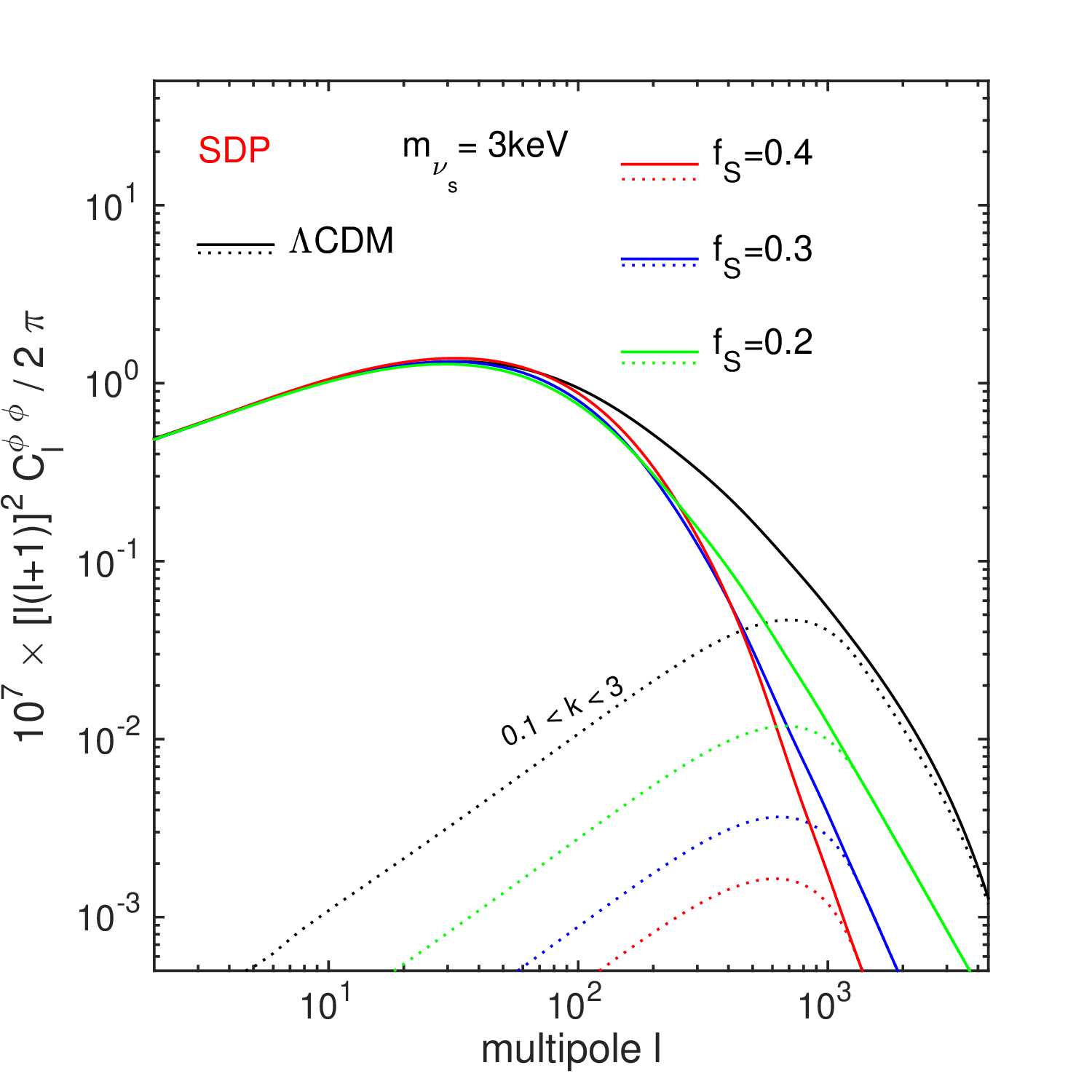}}%
\qquad
\subfigure{%
\includegraphics[width=7cm,height=7cm]{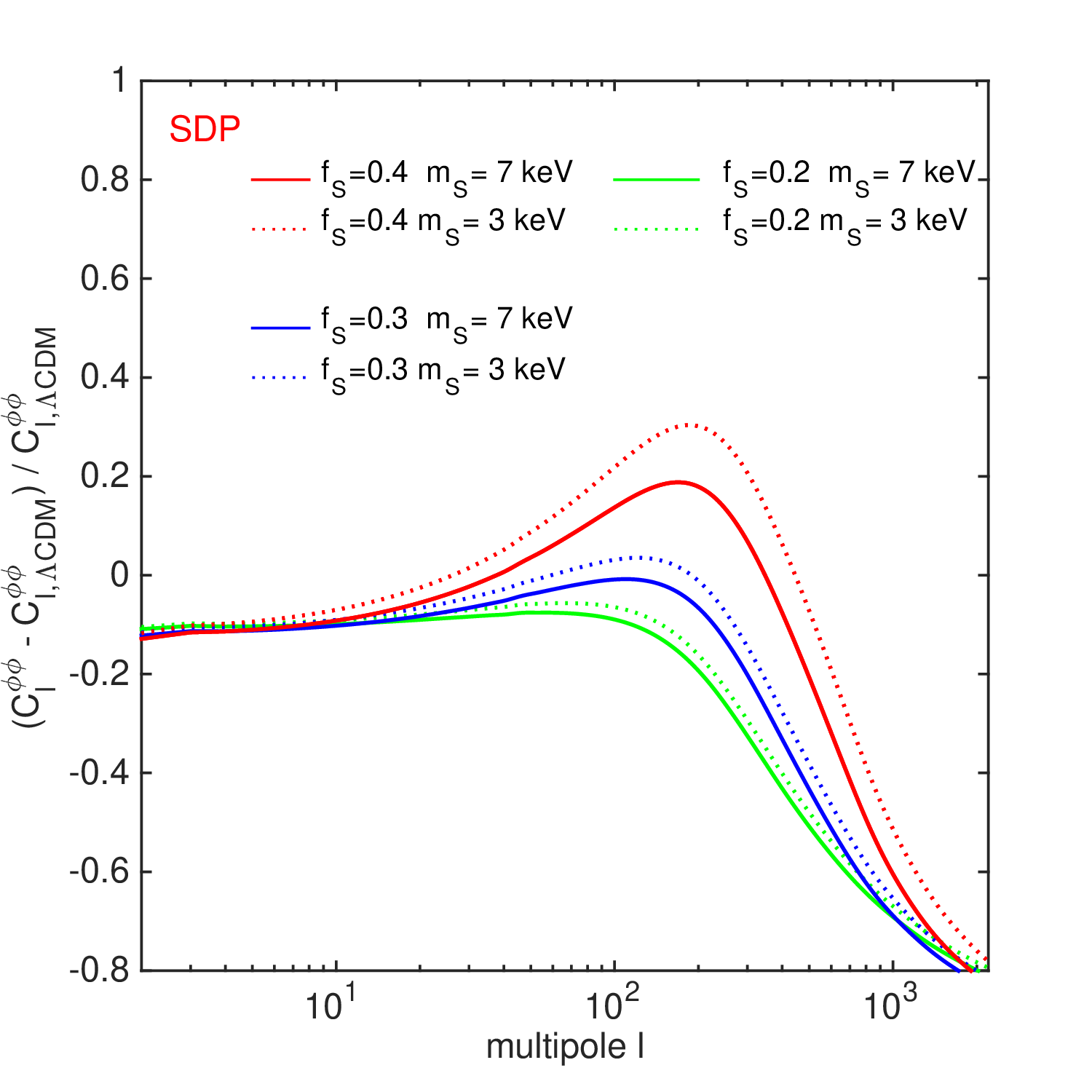}}%
\caption{{\it Left}:  The dependence of the deflection angle power spectra 
on sterile neutrino mass fraction $f_S$ in  models sharing 
the same sterile neutrino mass $m_{\nu_s}$, obtained in SDP scenario. 
For comparison we plot  the deflection angle power spectrum for $\Lambda$CDM model.
The contributions of the wave numbers in the range $0.1< k< 3$ Mpc$^{-1}$ are also indicated (dotted lines).
{\it Right}: Fractional differences between $C^{\phi \phi}_l$ obtained in SDP scenario
and in $\Lambda$CDM  model. The SDP mechanism parameters are indicated in Fig.~3. 
Other parameters are fixed to: 
$\Omega_b h^2=0.0226$, 
$\Omega_ch^2=0.112$, $\Omega_{\nu}h^2=0.00064$, $H_0$=~70~km~s~$^{-1}$~Mpc~$^{-1}$ and $\Omega_K=0$.}
\end{figure}
Depending on  both angular diameter distance and matter density fluctuations, 
$C^{\phi \phi}_l$  can break the degeneracy between  $m_{\nu_s}$ and $f_S$.
This can be seen explicitly in Fig.~5 where we show 
the dependences of $C^{\phi \phi}_l$ on $f_S$ for models sharing the same $m_{\nu_s}$ (left) and 
the fractional differences between $C^{\phi \phi}_l$ in $\Lambda$CDM  model and
in models shearing the same $f_S$ and different $m_{\nu_s}$ (right).\\
The observations of galaxy shear due to gravitational lensing can a achieve similar sensitivity at lower redshifts than the CMB gravitational lensing \cite{Aba2015}.

We conclude in Fig.~6 that {\sc Planck}+lens+BAO+DES datasets
have enough sensitivity to constrain the sterile neutrino mass and mass fraction inside the co-moving free-streaming horizon in both RP and SDP scenatrios.\\
\begin{figure}%
\centering
\subfigure{%
\includegraphics[width=7cm,height=6cm]{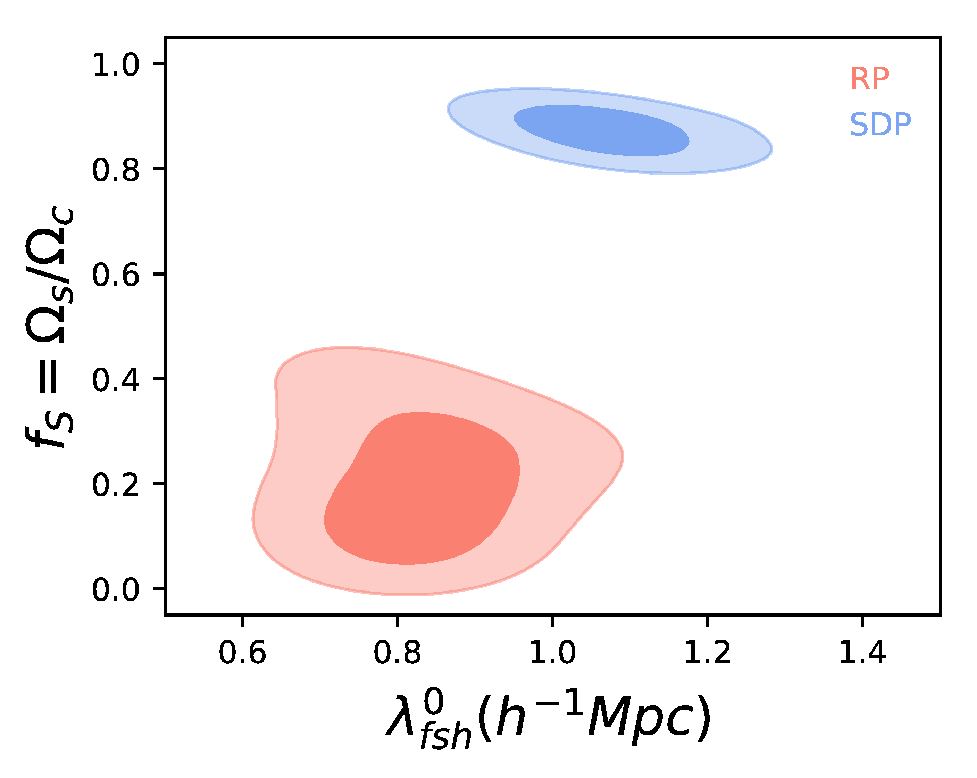}}%
\qquad
\subfigure{%
\includegraphics[width=7cm,height=6cm]{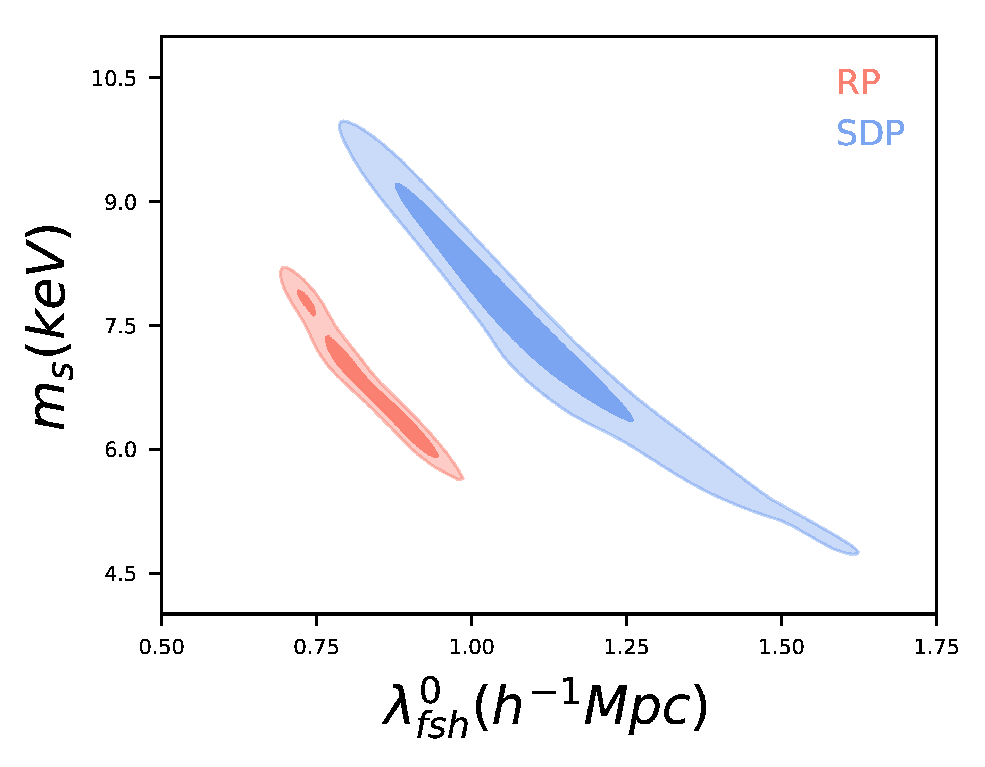}}%
\caption{{\it Left}: The marginalised likelihood  posterior distributions (at 68\% and 95\% C.L.) 
in $\lambda^0_{fsh}$ - $f_S$ plane  from the 
fit of RP and SDP models with {\sc Planck+lens+BAO+DES} datasets.
{\it Right}:  The same distributions 
in $\lambda^0_{fsh} - m_{\nu_s}$ plane.}
\end{figure}
\subsection{Constraints on sterile neutrino DM production parameters}
%==============================
\begin{figure}
\centering
\includegraphics[width=14cm,height=14cm]{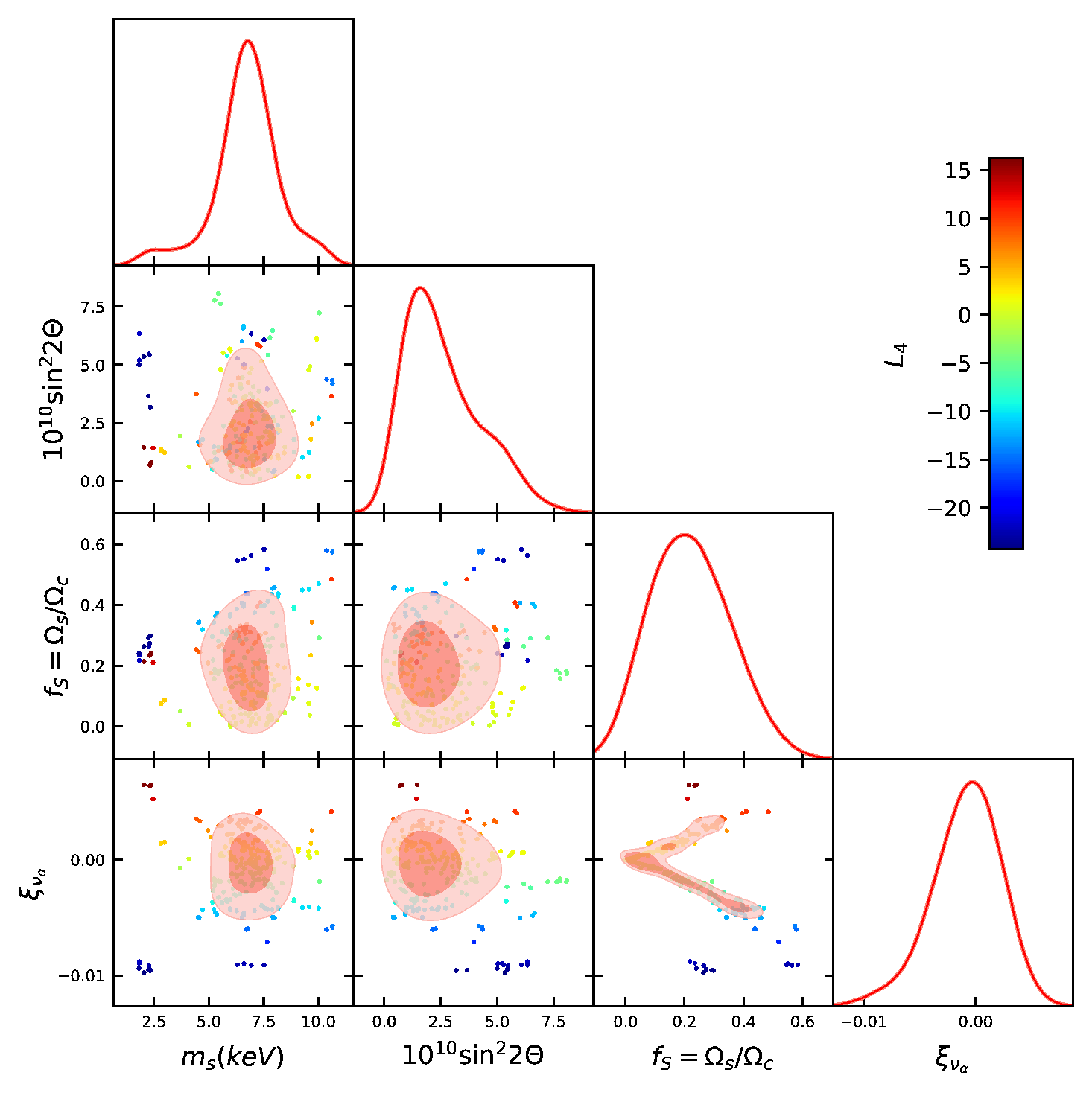}
\caption{The marginalised likelihood probability distributions 
and the joint confidence regions (68$\%$ and 95$\%$ CL)  for RP mechanism parameters 
colour-coded by the values of the initial lepton number $L_4=10^4 L_{\nu_{\alpha}}$. 
The  dominant effect on sterile neutrino RP production is given by  the lepton asymmetry $L_{\nu_{\alpha}}$, that 
sets the matter mixing angle $\sin^2 2 \theta$ to get a sterile neutrino mass, $m_{\nu_s}$. 
The best fit values of RP model parameters lead 
to $f_{S}$ =0.28$\pm$0.3 (68\% C.L.), 
indicating that RP  is a subdominant mechanism.
The sterile neutrino mass and  the mixing angle 
are in the parameter space of interest for DM decay interpretation of the ~3.5 keV X-ray line \cite{line}.}
\end{figure}
%===========================
{\it RP case:} Fig.~7 presents the likelihood probability distributions 
and the joint confidence regions obtained for the RP mechanism parameters. 
The dominant effect on sterile neutrino resonant production
is given by $\nu/{\bar \nu}$ chemical potential $\zeta_{\nu_{\alpha}}$, that sets 
the initial lepton number $L_{\nu_{\alpha}}$, which in turn sets the matter mixing angle $\sin^2 2 \theta$
to get the sterile neutrino mass  $m_{\nu_s}$. 
The best fit values of RP parameters lead to $f_S$ =0.28$\pm$0.3 (68\% C.L.), 
indicating that RP  is a subdominant mechanism. We find that  $m_{\nu_s}$ and  $\sin^2 2 \theta$ 
are in the parameter space of interest for  DM decay interpretation of the ~3.5 keV X-ray line
\cite{line}. \\
{\it SDP case:} Fig.~8 presents the likelihood probability distributions and the joint confidence regions obtained 
for the SDP mechanism parameters. 
The dominant effect on SDP production is given by the strength of the Higgs coupling, $\lambda_{H}$, that sets
$M_S$, and by the strength of Yukawa coupling, $y_{k}$, that sets the $m_{\nu_s}$.
The best fit values of SDP parameters lead to $f_{S}$~=0.86$\pm$0.07 (68\% C.L.), 
indicating that SDP  is a dominant mechanism.\\
Sterile neutrino mass predicted by RP and SDP mechanisms are consistent within 0.3$\sigma$
%==============================
\begin{figure}
\centering
\includegraphics[width=14cm,height=14cm]{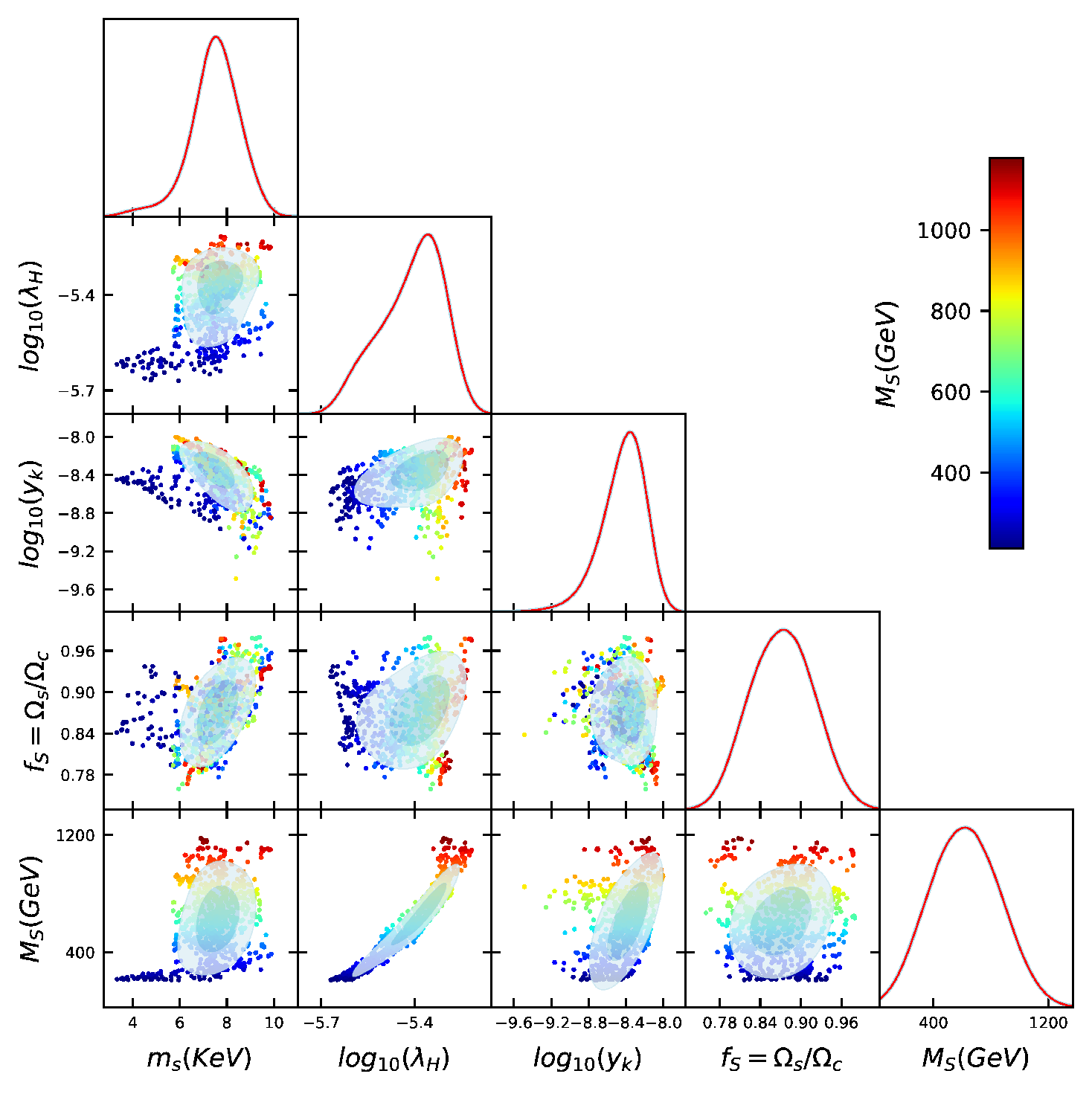}
\caption{The marginalised likelihood probability distributions 
and the joint confidence regions (68$\%$ and 95$\%$ CL)  for SDP mechanism parameters 
colour-coded by the scalar mass values $M_S$.
The dominant effect on SDP mechanism is given by the strength of the Higgs coupling, $\lambda_{H}$, that sets
$M_S$, and the strength of Yukawa coupling, $y_{k}$, that sets  $m_{\nu_s}$. 
The best fit values of the SDP parameters lead to $f_{S}$ =0.86$\pm$0.07 (68\% C.L.), 
indicating that SDP  is a dominant mechanism.
Sterile neutrino mass predicted by RP and SDP mechanisms are consistent within 0.3$\sigma$. }
\end{figure}

\subsection{Cosmological predictions of sterile neutrino production mechanisms}

\subsubsection{Acoustic scales}

%The characteristic angular size of CMB fluctuations, $\theta_s=r_s(z)/D_{A}(z)$, is determined by the
%co-moving sound horizon at recombination, $r_{s}(z)$, and the
%angular diameter distance at which fluctuations are observed, $D_{A}(z)$. 
%This is a quite robust parameter as its determination is based on observations when  CMB 
%fluctuations were entirely in the linear regime and therefore the higher order effects are negligible small. \\
%The transverse BAO angular scale, $\theta_d= r_{drag}(z) /D_A(z)$, where $r_{drag}(z)$
%is the comoving sound horizon at the end of the baryonic drag epoch, is the analog of $\theta_s$.\\
%The dependence on the unknown angular diameter distance 
%can be removed by taking the ratio $r_d/r_s=\theta_d/\theta_s$. 
As shown in the previous section, a tight constraint on the ratio $\theta_d/\theta_s$ implies a tight constraint on the radiation energy density at
photon decoupling, parametrised by  number of relativistic degrees of freedom, $N_{eff}$ \cite{Gr}.  
We find that the values of $N_{eff}$ obtained in RP and SDP scenarios are 
consistent with the SM value of $N_{eff}$
within 1.8$\sigma$ and 1.5$\sigma$ respectively.
Left panel from Fig.~9 shows that RP and SDP mechanisms are consistent within less 
than  1$\sigma$ in the $\theta_d/\theta_s$ - $N_{eff}$ plane.  \\
Motivated by the fact that $\Omega_m h^3$ is a well determined parameter orthogonal to the acoustic scale degeneracy in the flat cosmologies \cite{Percival,How}, we present in the right panel from Fig.~9 the confidence regions in $\theta_s$ - $\Omega_m h^3$  plane 
showing that RP and SDP mechanisms are also consistent within less 
than  1$\sigma$.
%==============================
\begin{figure}%
\centering
\subfigure{%
\includegraphics[width=6cm,height=5cm]{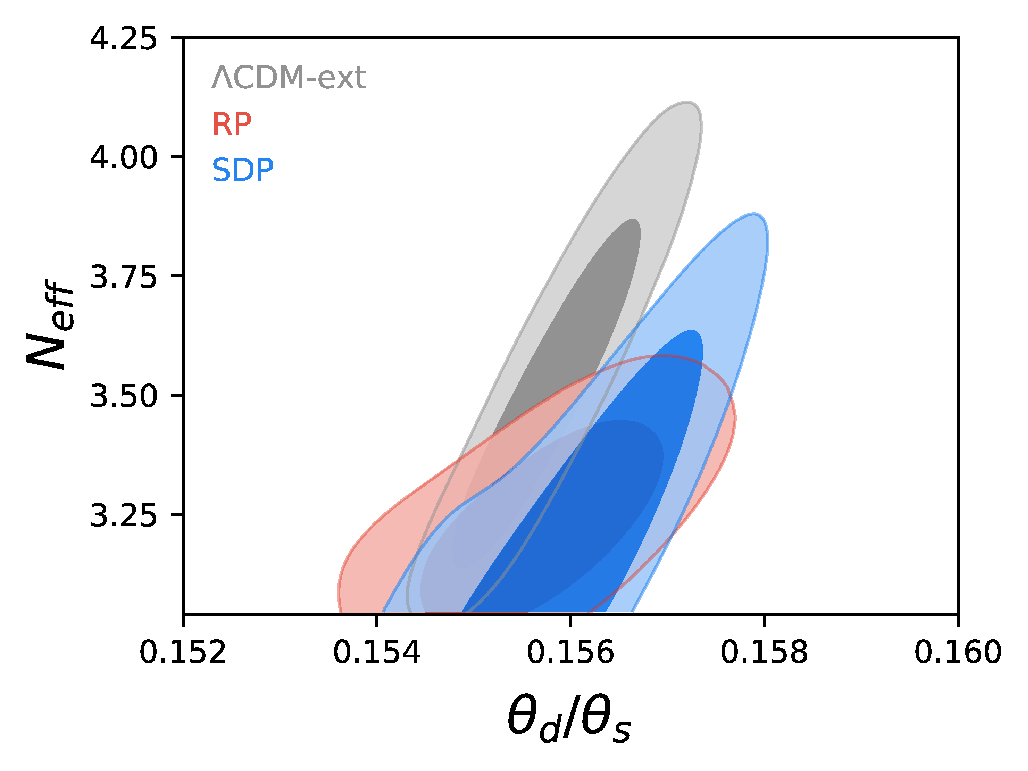}}%
\qquad
\subfigure{%
\includegraphics[width=6cm,height=5cm]{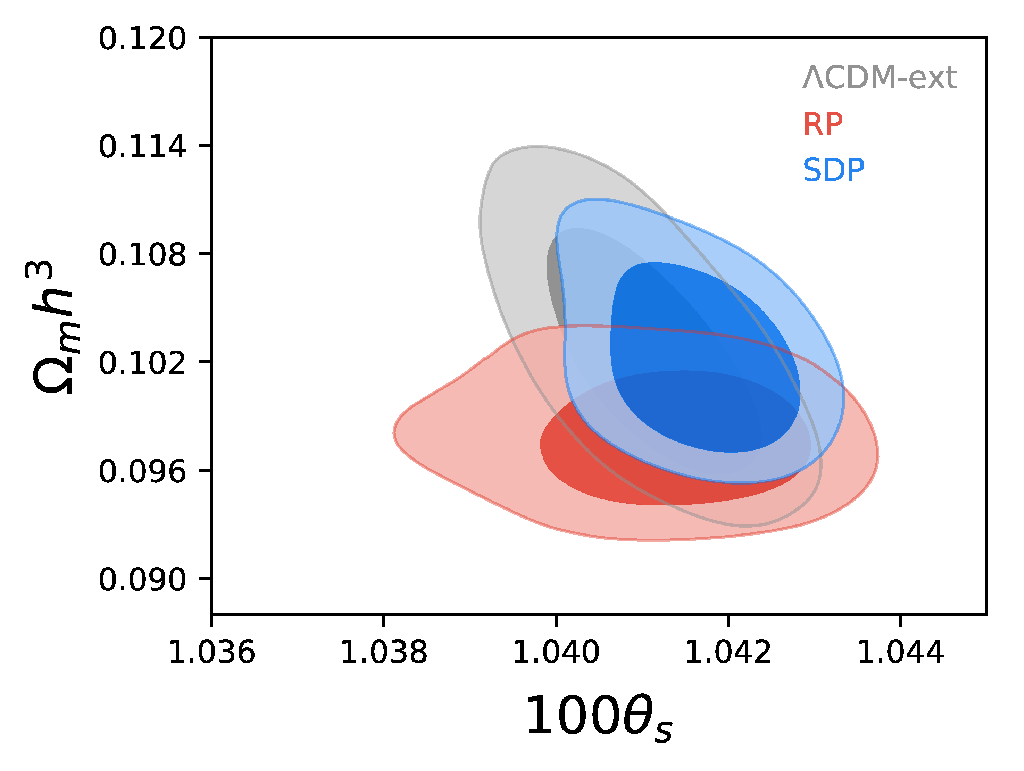}}%
\caption{The role of the acoustic scale measurements
to discriminate the sterile neutrino DM production mechanisms. The contours show the 68\% and 95\% C.L. 
{\it Left:} The confidence regions in  $\theta_d/\theta_s$ - $N_{eff}$ plane showing that
RP and SDP mechanisms are consistent within   less  than 1$\sigma$. 
{\it Right:} The confidence regions in $\theta_s$-$\Omega_mh^3$ plane
showing that RP and SDP mechanisms are also consistent within less 
than  1$\sigma$.}
\end{figure}
%=======================================
\subsubsection{Small-scale fluctuations}

The amplitude of the CMB acoustic Doppler peaks is exponentially suppressed  
on scales smaller than the Hubble radius at reionization due to the Thomson scattering of the 
free electrons produced at this epoch.
The amount of this suppression is given by $e^{-2\tau}$, where $\tau$ is the optical depth of the CMB photons.
{\sc Planck} high precision measurements of the CMB anisotropy  at small scales 
accurately constrain the damped amplitude while the CMB lensing potential power spectrum
provides the determination of the amplitude independent on the optical depth \cite{Planck13,Planck16}.
As the CMB power spectrum constraints 
the matter density fluctuations along the line of sight, 
the present-day  {\it rms} matter density power, $\sigma_8$,  is also determined. 
The CMB small-scale power fluctuations directly fixes the combination 
$\sigma_8 e^{- \tau}$ that is tightly constrained by the data \cite{Planck18}.\\
Also, the weak gravitation lensing of galaxies (cosmic shear) is sensitive to the matter fluctuations at small-scales, providing constraints on the combination $S_8 \equiv \sigma_8 (\Omega_m/0.3)^{0.5}$ \cite{DESc}.
%A number of studies used the  abundance of  
%Sunyaev-Zeldovich selected clusters \cite{Pl2016,Reichardt,Bleem,Anderson}  and of X-ray selected 
%clusters \cite{Mantz,Schi} to constrain $S_8$.\\
Fig.~10 illustrates the degree of consistency between the sterile neutrino RP and SDP mechanisms and the $\Lambda$CDM-ext model 
at small-scales. The left panel shows the impact of $\sigma_8 e^{-\tau}$ and $\Omega_m$. The RP and SDP models prefer 
higher values of $\sigma_8 e^{-\tau}$ that make them distinguishable from $\Lambda$CDM-ext  at 1.2$\sigma$ level. 
In the right panel of the same figure we show the impact of  $S_8\equiv \sigma_8(\Omega_m/0.3)^{0.5}$ and 
the Hubble parameter $H_0$. 
The  value of $S_8=0.792 \pm 0.024$ (68\% C.L.)  
obtained by DES survey from the combined clustering and lensing measurements \cite{DESc} 
is also indicated.
We find that $S_8$ values obtained in RP and SDP scenarios are consistent with that determined by DES 
within 0.6$\sigma$ and 1.5$\sigma$ respectively. 
%The preference of RP mechanism for smaller value of $H_0$ makes this mechanism  distinguible from SDP mechanism.
\begin{figure}%
\centering
\subfigure{%
\includegraphics[width=6cm,height=5cm]{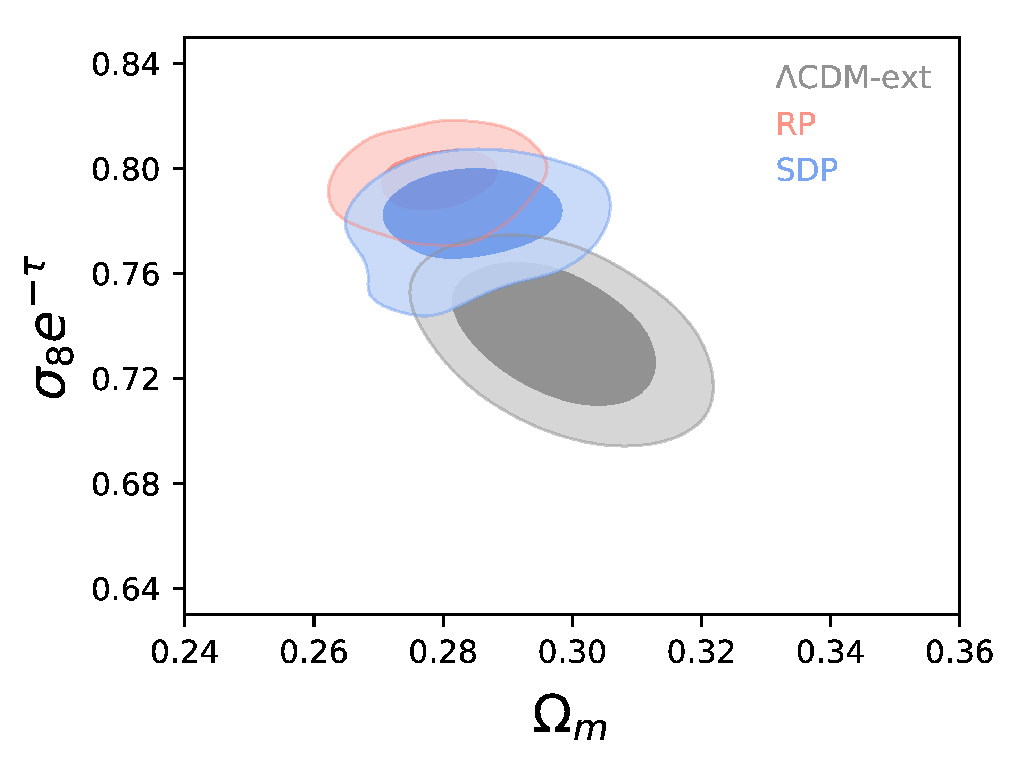}}%
\qquad
\subfigure{%
\includegraphics[width=6cm,height=5cm]{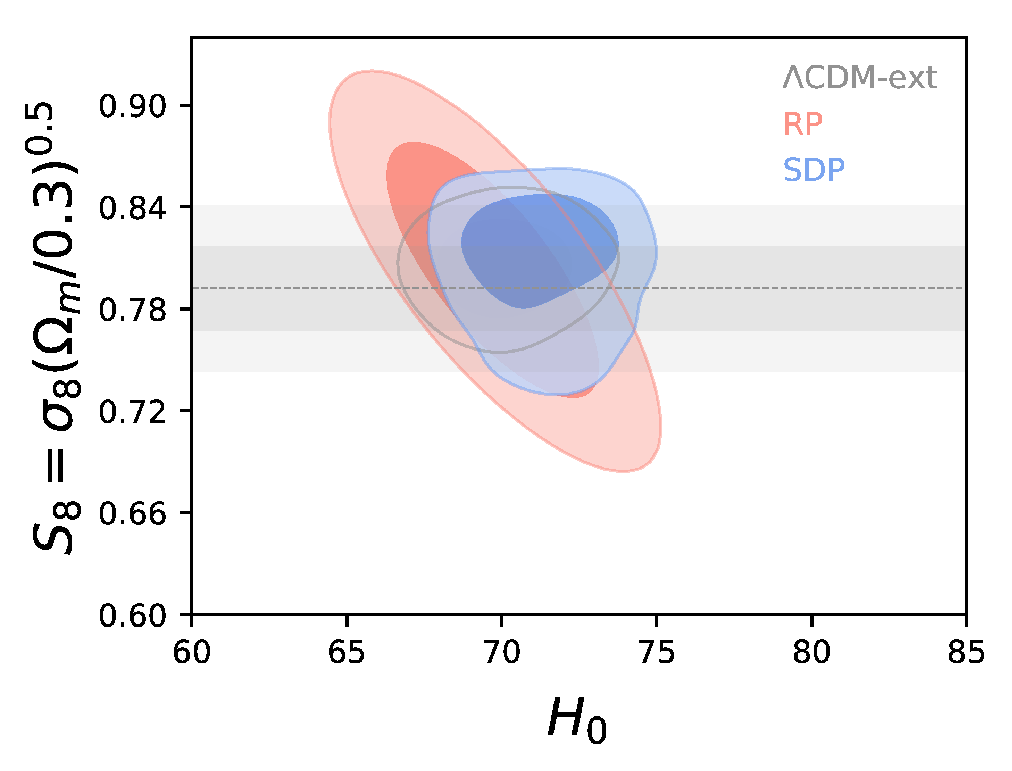}}%
\caption{The degree of consistency between sterile neutrino RP and SDP mechanisms and the $\Lambda$CDM-ext model 
at small-scales. The contours show the 68\% and 95\% C.L. \\
{\it Left}: The impact of $\sigma_8 e^{-\tau}$ and $\Omega_m$. 
The RP and SDP models prefer 
higher values of $\sigma_8 e^{-\tau}$ that make them distinguishable from $\Lambda$CDM-ext  model  at more than 1$\sigma$.
{\it Right}: The impact of  $S_8\equiv \sigma_8(\Omega_m/0.3)^{0.5}$ and 
Hubble parameter $H_0$.
The horizontal dashed line and the grey band mark the central value and $\pm 2\sigma$ 
error of $S_8$ value determined by DES survey from the combined clustering and lensing measurements \cite{DESc}.}
\end{figure}

\subsubsection{Low-redshift geometric probes}

We consider the low-redshift geometric probes, BAO and Hubble parameter $H_0$, to 
constrain the sterile neutrino DM production mechanisms. 
%Detected for the first time in the matter power spectrum by the 2dF Galaxy
%Redshift Survey \cite{Cole} and the SDSS redshift survey
%\cite{Eisenstein}, the BAO observations constrain the comoving size of the sound
%horizon $r_s(z_{drag})$  at the end of the drag epoch, $z_{drag}$, 
%when baryons and photons are completely decoupled. 
%The BAO surveys measure the acoustic-scale distance ratio: 
%\begin{equation}
%\label{BAO-Dv}
%d_z=\frac {r_s(z_{drag})} {D_V(z_{eff})},
%\end{equation}
%where $z_{eff}$ is the effective redshift for the population of observed galaxies and $D_V(z_{eff})$ is a combination of %the angular diameter distance $D_A(z)$, Hubble parameter $H(z)$ and 
%the speed of light $c$:
%\begin{equation}
%\label{BAO-Dv}
%D_V(z)=\left[ (1+z)^2 D^2_A(z)  \frac{cz} {H(z)} \right]^{1/3} \,.
%\end{equation}
We evaluate the characteristic BAO parameter (\ref{BAO-Dv}) at $z_{eff}=0.57$
reported by the BOSS DR11 survey \cite{Anderson14}.
Left panel from Fig.~11 presents constraints on our models in 
$H_0$ - $r_s/ D_V(z_{eff})$ plane. The horizontal dashed line and the grey bands mark the central value 
and $\pm 1\sigma$ and $\pm 2\sigma$ errors
of the BOSS measurement while the vertical dashed line and the grey bands do the same for 
$H_0$ determination from SHOES experiment \cite{H0-SHOES}. \\ 
On the other hand, the BAO features in the galaxy correlation function can be measured
in both line-of-sight and transverse directions, leading to joint constraints
on the angular diameter distance and the Hubble parameter at $z_{eff}$ \cite{Anderson14}.
Taking the fiducial sound horizon distance at the drag epoch $r_{fid}$=147.78 Mpc \cite{Alam17}, 
we compute the constraints on our models in  $D_A (z_{eff}) r_{fid}/r_{drag} - H_0(z_{eff}) r_{drag}/r_{fid}$ plane. 
The the join confidence regions are presented in the right panel from Fig.~11.\\
We conclude that present low-redshift geometric probes like BAO and $H_0$ start  to 
discriminate between the sterile neutrino RP and SDP mechanisms. 
However, the SDP scenario remains consistent with $\Lambda$CDM-ext model within  less than 1$\sigma$.
%===============================================
\begin{figure}%
\centering
\subfigure{%
\includegraphics[width=6cm,height=5cm]{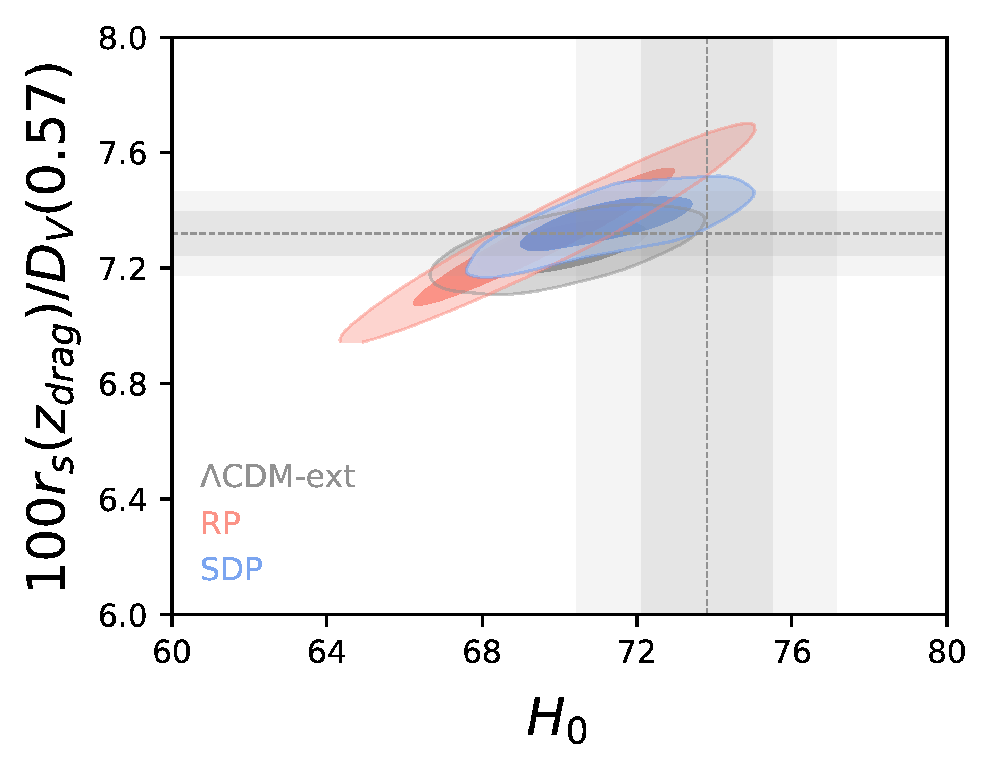}}%
\subfigure{%
\includegraphics[width=6cm,height=5cm]{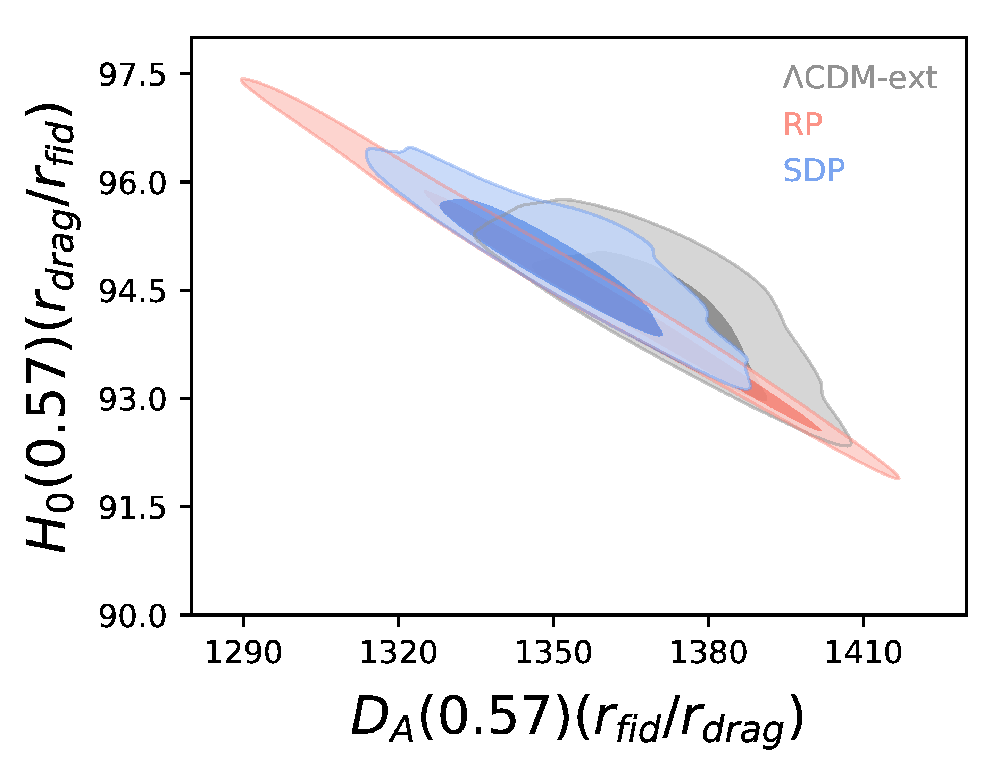}}%
\caption{ The role of the low-redshift geometric probes 
to discriminate between the sterile neutrino DM production mechanisms. The contours show the 68\% and 95\% C.L. \\
{\it Left}: The impact of Hubble parameter $H_0$ and of the BAO characteristic parameter $r_s/D_V(z_{eff})$. 
The horizontal dashed line and the grey bands mark the central value 
and $\pm 1\sigma$ and $\pm 2\sigma$ errors
of the BOSS measurement at $z_{eff}=0.57$ \cite{Anderson14}, while the vertical dashed line and the grey bands do the same for 
$H_0$ determination from SHOES experiment \cite{H0-SHOES}.
{\it Right}: The role of BAO measurements 
on line-of-sight and transverse directions, leading to joint constraints of 
$D_A (z_{eff})/r_{drag}$ and $H_0(z_{eff}) r_{drag}$.  We take the fiducial sound horizon distance at the drag epoch  $r_{fid}$=147.78 Mpc
\cite{Alam17}.}
\end{figure}
%===========================%====================================
\section{Conclusions}

\begin{figure}
\centering
\includegraphics[width=16cm,height=9cm]{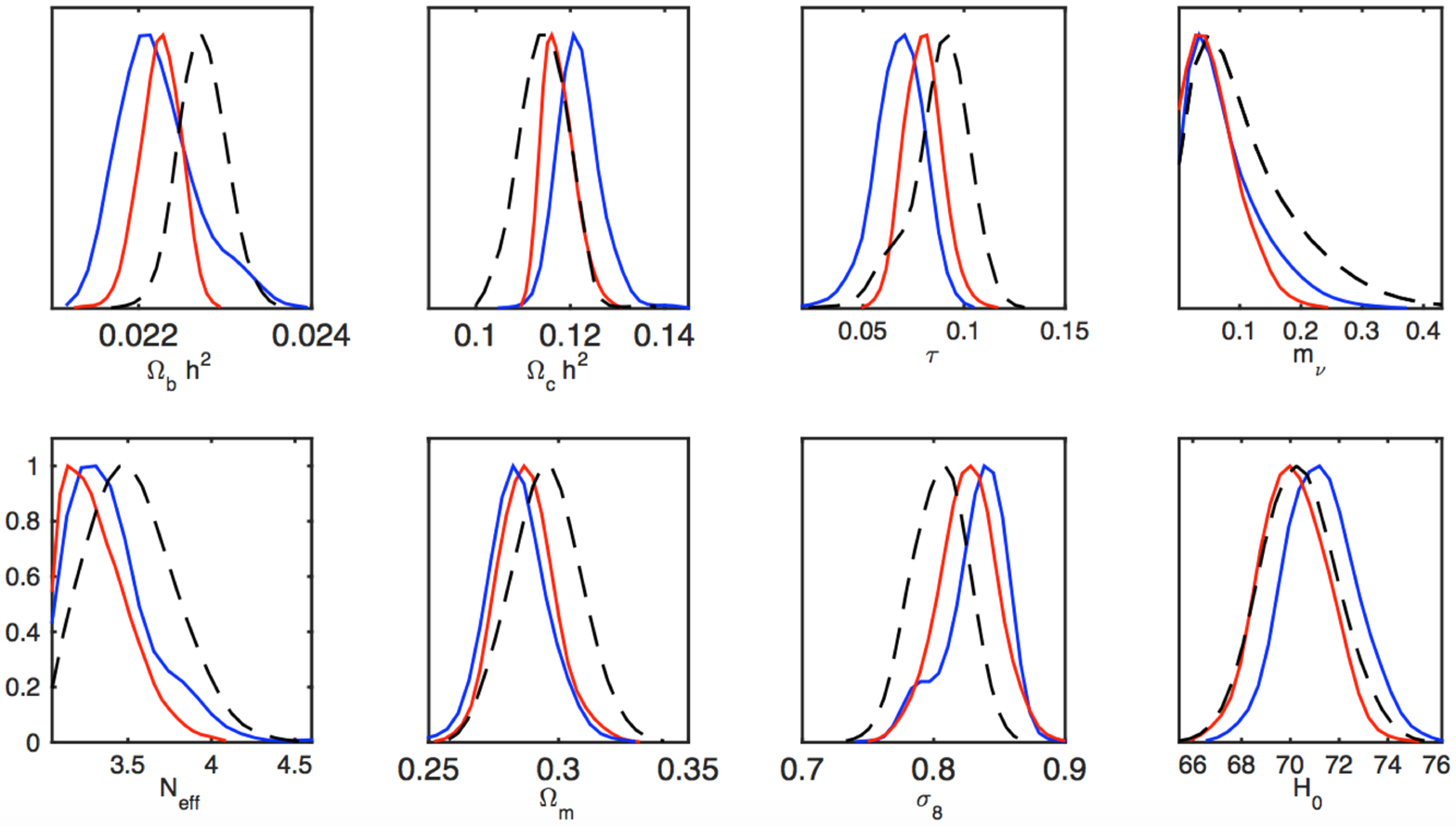}
\caption{ The marginalised  likelihood probability distributions obtained from the fit with 
{\sc Planck}+lens+BAO+DES  datasets of the $\Lambda$CDM-ext model (black dashed lines), 
RP model (red lines) and SDP model (blue lines).}
\end{figure}

In this paper we place constraints on sterile neutrino RP and  SDP  mechanisms  assuming
that sterile neutrino represents a fraction $f_S$ from the total CDM energy density. \\
So far, the  keV sterile neutrino properties 
have been addressed   under the 
assumption that sterile neutrinos are all of the DM,
by evaluating their impact on the co-moving free streaming horizon 
that  relates on the average velocity distribution. 
For  such models, characterised by highly non-thermal momentum distributions, the average momentum is subject of uncertainties, leading to the fail of free-streaming horizon 
in constraining their parameters. 

For our cosmological analysis, we complement the CMB anisotropies measurements with CMB lensing gravitational potential measurements, that are
sensitive to the DM  distribution  out to high redshifts
and with the cosmic shear data, that constraints the gravitational potential at lower redshifts than CMB. We also use the most recent 
low-redshift BAO measurements that are insensitive  to the non-linear effects,
providing robust geometrical tests.

We show that for models sharing the same $f_S$, the accurate determination of the acoustic scale from CMB anisotropy measurements
breaks the degeneracy of Hubble parameter at the photon decoupling, constraining $m_{\nu_s}$ in RP scenario,
while the BAO measurements constrain $m_{\nu_s}$ at lower redshifts.
We evaluate the co-moving free-streaming horizon for RP and SDP models showing that,
the deflection angle power spectrum of the CMB lensing potential, $C^{\phi phi}_l$ is sensitive to both sterile neutrino
production mechanisms for the multipole range involved in this analysis (40 $\leq l \leq$ 400) with the increased 
wave-number of power suppression in RP case. Depending on both angular diameter distance and matter density fluctuations, we show that $C^{\phi phi }_l$ can break the degeneracy between $f_{S}$ and $m_{\nu_s}$
in our models.
We also show that our datasets have enough sensitivity to constrain 
the sterile neutrino mass and mass fraction inside the co-moving free-streaming 
horizon in both RP and SDP scenarios. \\
The best fit parameters obtained from our cosmological analysis are presented in Tab.~3 and Fig.~12.
For RP case we find that the best fit values of $m_{\nu_s}$ and $\sin^2 \theta$  
are in the parameter space of interest for sterile neutrino DM decay interpretation of the 3.5 keV X-ray line
with a DM mass fraction $f_{S}=0.28 \pm 0.3$  (at 68\% CL) that excludes  the assumption of
sterile neutrinos as being all of the DM.  For SDP case we find $f_{S}=0.86 \pm 0.07$  (at 68\% CL), 
in agreement with the upeer limit constraint on $f_S$ from the X-ray non-detection and Ly-$\alpha$  
forest measurements that rejects $f_S=1$  at 3$\, \sigma$ level \cite{Slosar}. \\
The sterile neutrino mass predicted by both RP and SDP models are consistent within 0.3$\sigma$.\\
%can be compensated by changes in cosmological parameters. \\
We analysed the possibility to distinguish  between  RP and SDP scenarios through their impact on the
acoustic scales, the small scale fluctuations and the low-redshift geometric observables, obtaining 
cosmological constrains that clearly show that  
the present-day cosmological data start to discriminate between different 
sterile neutrino DM production mechanisms. \\
However, we expect 
the future BAO and weak lensing surveys, such as
{\sc EUCLID}, to provide much robust constraints.\\
\begin{table}
\caption{The table shows the mean values and the absolute errors of the main parameters
obtained from the fit of $\Lambda$CDM-ext, RP and SDP models with the data-sets. The errors are quoted at 68\% C.L. 
The upper limits are quoted at  95\%C.L.
The first group of parameters are the base cosmological parameters sampled in the Monte-Carlo Markov Chains 
analysis with uniform priors.
The others are derived parameters.}
\begin{center}
\begin{tabular}{lccc}
\hline \hline
                 & $\Lambda$CDM-ext   &        RP     &      SDP\\
Parameter&              &                  &             \\
\hline
$\Omega_b h^2$& 0.0223$\pm$0.0002& 0.0222$\pm$  0.0003&0.0219$\pm$ 0.0003 \\
$\Omega_c h^2$&0.122$\pm$0.004&0.118$\pm$0.003&  0.121$\pm$0.004\\
100$\theta_{MC}$&1.0412$\pm$0.0008& 1.0404$\pm$0.0011 & 1.0413$\pm$0.0009 \\
$\tau$&0.087$\pm$0.015 &0.079$\pm$ 0.009&0.069$\pm$0.012\\
$\sum m_{\nu}$& $<$ 0.321&$<$0.249& $<$ 0.198\\
$N_{eff}$& 3.520 $\pm$0.259 &3.313$\pm$0.109& 3.380 $\pm$0.243 \\
$f_{S}$&               & 0.281$\pm$0.03  &0.860 $\pm$0.071 \\
$\sin^2 2\theta$ &   & 2.460 $ \pm$  1.750 \\
$10^{3} \zeta_{\nu_{\alpha}}$&  & -0.822$\pm$ 2.691 & \\
M$_S$ (GeV)&  &   & 533.60 $\pm$ 47.21 \\ 
$10^{-6} \lambda_H$&  &    &  3.780$\pm$0.642     \\
$10^{-9}y_k$& & & 3.451$\pm$ 1.820\\
\hline
$\Omega_m$&0.295$\pm$0.013&0.287$\pm$0.011&0.284$\pm$0.011\\
$\sigma_8$&0.808$\pm$0.021&0.801$\pm$0.022 &0.832$\pm$0.019         \\
$H_0$&   70.512$\pm$1.556& 70.142$\pm$1.355&71.210$\pm$1.433 \\
$m_{\nu_{s}}$(keV)&     &  6.831 $\pm$1.630    &  7.882 $\pm$  0.731    \\
$L_4=10^4 L_{\nu_{\alpha}}$  &  & -2.081 $\pm$ 6.710& \\
$100 \theta_s$& 1.0411$\pm$0.0002& 1.0391$\pm$ 0.0011 &1.0421 $\pm$ 0.0011\\
$100 \theta_d$&0.1622$\pm$  0.0002&0.1619$\pm$0.0061&0.1632$\pm$0.0011 \\
\hline \hline
\end{tabular}
\end{center}
\end{table}
\begin{acknowledgments}

This work was partially supported by UEFISCDI Contract 18PCCDI/2018. \\
We also acknowledge the use of the GRID computing system facility at the Institute of Space Science Bucharest and would like to thank the staff working there.%The author would like to thank to Rose Lerner and Dmitry Gorbunov for helpful discussions.\\
%This work  was partially supported by CNCSIS Contract 539/2009 and by
%ESA/PECS Contract C98051.
\end{acknowledgments}

\end{document}